\documentclass[sigconf,nonacm]{acmart}

\usepackage{algorithmic}
\usepackage{graphicx}
\usepackage{textcomp}
\usepackage{xcolor}
\usepackage{booktabs}
\usepackage{microtype}
\usepackage{xspace}
\usepackage{float}
\usepackage{fontawesome}
\usepackage{subcaption}
\PassOptionsToPackage{hyphens}{url}\usepackage{hyperref}
\usepackage{pifont} 
\usepackage[capitalise]{cleveref}
\usepackage{enumitem}

\newcommand{\fakeparagraph}[1]{{\vskip 2pt \noindent\bfseries #1.}}
\newcommand{\xmark}{\ding{55}}%
\newcommand{\cmark}{\ding{51}}%
\newcommand{\specialCat}{\ding{105}}%
\newcommand{\eg}{e.g.,\xspace}
\newcommand{\ie}{i.e.,\xspace}
\newcommand{\etAl}{et\,al.\xspace}

\microtypecontext{spacing=nonfrench}
\begin{document}

\title{From Third-Party to First-Party: Measuring and Protecting Against Modern Web Tracking Mechanisms}

\author{Christian B\"ottger}
\orcid{0009-0006-3467-9965}
\email{boettger@internet-sicherheit.de}
\affiliation{%
  \department{Institute for Internet Security}
  \institution{Westphalian University of Applied Science}
  \country{}
}
\affiliation{%
  \institution{Doctoral School NRW}
  \department{Department of Computer and Data Science}
  \country{}
}

\author{Tareq Khouja}
\orcid{0009-0004-0470-5865}
\email{khouja@internet-sicherheit.de}
\affiliation{%
  \department{Institute for Internet Security}
  \institution{Westphalian University of Applied Science}
  \country{}
}

\author{Norbert Pohlmann}
\orcid{0009-0007-6221-7327}
\email{pohlmann@internet-sicherheit.de}
\affiliation{%
  \department{Institute for Internet Security}
  \institution{Westphalian University of Applied Science}
  \country{}
}

\author{Nurullah Demir}
\orcid{0000-0001-8721-4412}
\email{nurullah@cs.stanford.edu}
\affiliation{%
  \institution{Stanford University}
  \country{}
}

\author{Tobias Urban}
\orcid{0000-0003-0908-0038}
\email{urban@internet-sicherheit.de}
\affiliation{%
  \department{Institute for Internet Security}
  \institution{Westphalian University of Applied Science}
  \country{}
}

\begin{abstract}
Web user tracking has always been a cat-and-mouse game between privacy-conscious users and trackers. Recently, this conflict has driven a shift from third-party tracking toward first-party tracking (FPT) and server-side tracking (SST). 
By relocating tracking logic to the browser's first-party context or the website's backend, these mechanisms obscure data flows and render traditional client-side detection tools increasingly ineffective. 
Despite the growing adoption of these techniques, our understanding of their deployment at scale remains limited, and generalized protection mechanisms are lacking.

In this work, we conduct a large-scale measurement of top sites to assess this shift and the prevalence of FPT and SST. We develop a provider-independent methodology to detect these mechanisms and find that over 54\% of analyzed sites now deploy FPT or SST-related techniques. By clustering scripts based on their similarity and constructing a network graph, we demonstrate that the ecosystem is densely connected and dominated by major vendors like Google. 
Finally, we demonstrate that current filter lists are largely ineffective against first-party tracking, and we propose new rules to address this gap. We show that these rules block 63\% more requests than traditional filter lists. 
\end{abstract}

\maketitle

\section{Introduction}
\label{sec:introduction}

On the modern Web, targeted advertising (ads) is the primary revenue model for websites~\cite{iab.2024}, and behavioral analytics are important for optimizing websites~\cite{JARVINEN2015117}, both of which rely on user tracking.
Consequently, providers employ pervasive tracking techniques to collect extensive user data.
These practices are often viewed as privacy-invasive, as they often occur without proper consent~\cite{utz_consent_2019,nguyen_consent_2022} and offer limited transparency mechanisms to meaningfully inform users~\cite{urban.2019,Kretschmer.2021,Prince.2024}. %
Traditionally, third-party cookies have served as the predominant mechanism for this data collection~\cite{urban_adnetworks_2020,Acar.tracking.2014,gonzalez_cookies_2017,10.1145/2736277.2741679}.
A recent direction is that trackers implement tracking methods that do \emph{not} rely on third-party cookies~\cite{WebAlmanac.2024.Privacy} by moving the tracking mechanisms into the website's backend or into the first-party context. 
This technique is called \emph{first-party tracking} (FPT)~\cite{demir.2022} or \emph{server-side tracking} (SST)~\cite{google_server_side_tagging,elfraihi:hal-04665102,Alegri2025}.
In contrast to traditional client-side tracking, FPT moves data collection and transmission from the user's browser to the website's backend %
or the first-party context of a page~\cite{cookiegraph}. 
In both cases, website providers host tracking mechanisms in a first-party context or use third-party resources that can access the first-party context, enabling them to evade privacy-enhancing tools and making URL-based detection challenging. Furthermore, this relocation obscures the data flow, as users cannot easily identify third-party tracking-related requests or the entities involved~\cite{Fouad2024}. %

While prior studies have offered insights into specific implementations, such as Meta or Google Tag Manager~\cite{elfraihi:hal-04665102, Alegri2025}, or addressed particular legal~\cite{Fouad2024} and technical challenges~\cite{cookiegraph}, the broader ecosystem remains less explored.
To bridge this gap, we develop a generic methodology to identify FPT and SST
and conduct large-scale measurements across 13k sites and 750k pages to assess their adoption.
We assess which parties are involved in this new tracking ecosystem and test if similar tracking code is in first- and third-party tracking methods.
Furthermore, since blocking FPT requests is challenging, we developed a statistical method to generate blocking rules compatible with ad blockers, thereby protecting users from FPT and SST.

We find that over 54\% of analyzed sites have adopted FPT or SST. By clustering the JavaScript involved in these techniques, we identify an ecosystem operated by 492 distinct entities, yet highly centralized around a few major vendors (\eg Google).
Notably, we find that a significant number of third-party trackers have begun operating in first-party contexts, accounting for over 8\% of the analyzed tracking requests.
To address the limited efficiency of block lists, we derive 181 FPT-blocking rules using a statistical approach, showing that they block 63\% more first-party tracking requests than existing lists with a minor impact on page breakage.

\smallskip
\noindent
In summary, we make the following contributions:
\begin{itemize}[leftmargin=*]
    \item \textbf{Prevalence of FPT and SST.} We show that, with 13,187 sites using cookies potentially linked to FPT or SST, these technologies are an emerging and prevalent privacy threat on the Web. Further, we show that trackers have begun migrating their activities into the first-party context, as we find similar scripts in both contexts.

    \item \textbf{First-party tracking ecosystem analysis}. By clustering tracking scripts into 492 entities, we reveal a centralized ecosystem dominated by major vendors (\eg Google), demonstrating that first-party tracking largely relies on existing third-party infrastructure.   
    
    \item \textbf{First-party-tracking protection mechanisms.} We design and evaluate a URL-based detection and mitigation pipeline that utilizes FPT-specific query-parameter patterns and translates them into blocking rules. We show that the generated rules block notably more FPT-related requests (63\%) than common blocklists without causing considerable page breakage. 
\end{itemize}

\section{Background}
\label{sec:background}
Here, we introduce the terminology used in this work and describe the basis of first-party and server-side tracking.

\subsection{Terminology}
The term \emph{site} refers to the registrable portion of a domain, commonly known as the ``effective Top-Level Domain plus one'' (eTLD+1)~\cite{Urban.WWW.2020,Calzavara.Headers.2021,boettger.2025,Demir.updates.2021,PublicSuffixList.2025}.
For example, in \nolinkurl{https://www.example.edu/}, the eTLD+1 is \nolinkurl{example.edu}. 
In contrast to a site, a \emph{page} (or \emph{webpage}) is a specific URL and the document accessible at that address. Thus, a site can have multiple pages that contain the site's content.

Cookies are small key–value pairs stored by the browser to persist state across requests~\cite{rfc6265,Urban.WWW.2020}. We refer to cookies by their setting context (e.g., first-party cookies). Notably, third-party scripts can set first-party cookies as long as they originate from the visited domain~\cite{rfc6265}.

\subsection{Server-Side and First-Party Tracking}
\label{sec:sst}
\begin{figure}[tb]
    \centering
    \begin{subfigure}{0.45\columnwidth}
        \includegraphics[width=\columnwidth]{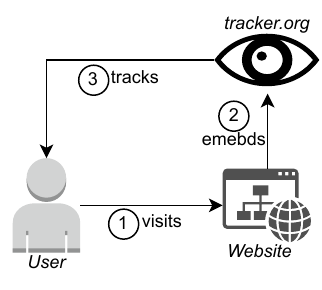}
        \caption{Client-side tracking.} %
        \label{fig:cst}
    \end{subfigure}
    \hfill
    \begin{subfigure}{0.45\columnwidth}
        \includegraphics[width=\columnwidth]{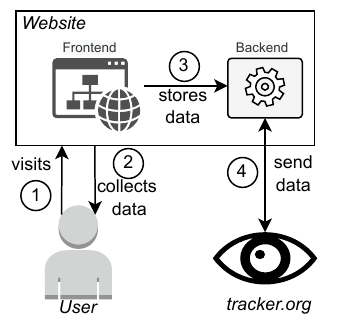}
        \caption{Server-side tracking.} %
        \label{fig:sst}
    \end{subfigure}     
    \caption{Comparison of client-side and server-side tracking.}
    \label{fig:cst_vs_sst}
\end{figure}

 \emph{Server-Side Tracking} (SST) and \emph{First-Party Tracking} (FPT) shift tracking activities from third parties to the website's backend or the user's first-party context, introducing distinct challenges for transparency and detection.
As more and more defense techniques against client-side tracking are employed, providers of tracking technology need to adapt their data-collection practices. Consequently, major tracking providers such as Google, TikTok, or Meta have introduced a technique referred to as server-side tracking ~\cite{Google2020ServerSideTagging,Meta2025SignalsGateway,Jentis2024Whitepaper}.
SST moves essential parts used for tracking into the backend of the visited site (\ie the first party). 
The server collects or aggregates the data of interest and, after preprocessing, forwards it to the tracker~\cite{fouad2020,kollnig2021before}.
Thus, a key difference from client-side tracking is that this type of tracking can occur without the client being aware of any communication with a third party.
\Cref{fig:cst_vs_sst} shows a high-level comparison between both techniques. 
While the name ``server-side'' tracking implies that tracking \emph{only} happens on the server, current mechanisms rely on storing an identifier on the client in the form of first-party cookies~\cite{cookiegraph}.
The client-side component of an SST deployment is minimal, a persistent identifier is stored in a first-party cookie, which the first-party server reads on each visit and transmits to the tracker as part of its backend processing~\cite{elfraihi:hal-04665102, google_sst_fundamentals, google_server_side_tagging, tiktok_gtm_pixel_2025}. From the browser's perspective, this tracking can happen without any request to a tracker domain.
This architectural property is one reason SST is harder to detect and protect against: the first-party domain varies by definition across sites and cannot be part of a static block list~\cite{Fouad2024}.

Further, first-party tracking (FPT) refers to a tracking model in which the user identifier is stored in the first-party context (\eg in a first-party cookie).
Because the identifier is set and accessed in a first-party context, it is not subject to many browsers' defenses that restrict third-party cookies. 
This can be achieved by embedding tracking logic directly into the website's first-party JavaScript code, by proxying third-party scripts through a first-party subdomain via DNS reconfiguration~\cite{9403411}, or by configuring a tag management system such as Google Tag Manager to deliver tracking code within the first-party context~\cite{Alegri2025}. Because the browser classifies these resources as first-party, neither third-party cookie restrictions nor domain-based blocking rules apply. The tracking identifier is stored as a first-party cookie via the ``document.cookie'' API~\cite{rfc6265} and persists across browsing sessions.
Hence, first-party tracking enables cross-site or long-term user tracking while bypassing privacy mechanisms designed to limit traditional client-side tracking (\eg third-party tracking)~\cite{demir.2022}.

While SST and FPT operate differently, both rely on first-party cookies as persistent user identifiers. We focus on detecting and analyzing these cookies, and henceforth use FPT to refer to both mechanisms.

\section{FPT Measurement Framework}
\label{sec:sst_measurement}
In this section, we detail our data collection, measurement, preprocessing, and analysis. %
Artifacts of our work are provided in a code repository (see \cref{sec:data_sharing}).

\subsection{Data Collection}
\label{sec:data_collection}
The scope of this study is to assess client-side indicators FPT. 
We do not aim to capture communication between servers and, therefore, cannot draw any conclusions about what is happening on the server side. Yet, we argue that if we find client-side first-party indications for user tracking, it is processed by a server for that purpose (\eg we assume that if on the client side FPT-related tracking code is executed, then on the server side a corresponding component exists).
To understand the FPT tracking ecosystem and develop appropriate countermeasures, we need to collect a representative sample of FPT-related first-party cookies (see \cref{sec:sst}), the parties that set them, and the websites that use such tracking techniques.
To collect this data, we use the MultiCrawl Web measurement framework~\cite{Demir_MultiCrawl}, which is based on OpenWPM~\cite{englehardt2016online}. %
During a measurement run, we instrument the framework to record HTTP requests and responses (including headers), cookies set via HTTP headers, JavaScript interaction with cookies (\eg we instrumented the \texttt{content\_script\_instrument} and \texttt{save\_content} function), and download all JavaScript resources (files).
OpenWPM uses the Firefox browser, which we configure to prevent third-party cookies from being set or sent in third-party contexts. 
Thus, websites might resort to first-party tracking, as third-party cookie-based tracking is not possible.
Further, to minimize detection by anti-bot mechanisms, we applied best practices to hide that OpenWPM is in use~\cite {krumnow2022}. %

To increase the number of cookies in our dataset, we interact with the consent banners on webpages.
We use the browser extension \emph{Consent-O-Matic}~\cite{consentomatic2021} and modify it to accept all cookies. %
We choose Consent-O-Matic because it is a major extension for this use and has been validated as effective in previous work~\cite{Demir.banner.24,Nouwens2022}.
For our measurement, we use the Tranco\footnote{\url{https://tranco-list.eu/list/QG4V4/1000000}} list~\cite{LePochat2019} generated on  April 2nd, 2025. %
We use the top 5k sites from the list and randomly select 5k sites from each of the following buckets: 5,001-10k, 10,001-50k, 50,001-250k, 250,001-500k~\cite{demir_2023}, resulting in a total of 25k sites. 
For each of the 25k sites, we visit 25 pages~\cite{Urban.WWW.2020}.

Overall, we run four measurements simultaneously on different virtual machines (VMs) from two locations (\ie Germany and the United States). 
Our framework follows best practices for large-scale Web measurements~\cite{demir_2023,Demir_rep_2022} (\eg multiple crawls and multiple locations) to capture a realistic, representative dataset on the prevalence of server-side tracking in the field. 
We chose to run the experiments in the U.S. because sites typically use more cookies when users visit them from the U.S. rather than from the EU~\cite{Demir.banner.24}. We performed two measurements from each location to increase the validity of the results and to robustly identify first-party tracking cookies. %
All of the VMs are located in Germany. We connected all machines via a VPN to two different locations: two are connected to VPN gateways in the U.S., and two machines are connected to VPN gateways in Germany. We use \emph{NordVPN}~\cite{nordvpn} as a VPN provider.%
VPN providers may impact measurements (\eg by inserting ads)~\cite{Khan.2018}. However, we found no evidence of this for NordVPN. We deactivated NordVPN's integrated AdBlock mechanism. Sites detecting VPN use would affect all measurement profiles equally.

\subsection{Data Preprocessing}
\label{sec:data_processing}

\fakeparagraph{Identifying Tracking Cookies}
\label{sec:cookies_identify}
In this work, we want to assess the impact of first-party tracking on the tracking ecosystem.
Thus, we need to determine whether first-party cookies are used for tracking~\cite{cookiegraph}.
One significant challenge in this task is that first-party cookies are often used to identify sessions and maintain state within the otherwise stateless HTTP protocol.
Such session identifiers are similar to tracking IDs, as they serve a comparable purpose: identifying the same user across page visits and HTTP requests.
Session IDs typically expire when a browser is closed, whereas user identifiers are designed to persist across multiple sessions. %

To select an appropriate identification method, we assess three prior approaches. The first is a community-driven approach that uses classifications from Cookiepedia~\cite{Cookiepedia.2024}, which curates a mapping of cookie names to their potential purposes.
To assess its suitability, we applied the Cookiepedia database to our collected first-party cookies. Cookiepedia classified only 3,914 (0.09\%) of the 477,231 first-party cookies as tracking cookies, confirming the limited coverage of community-driven databases for this emerging phenomenon. In contrast, our heuristic (described below) identified 0.14\% of cookies as tracking-related and discovered 16,881 (51\%) additional cookies that Cookiepedia could not classify. The heuristic missed only 1 cookie that Cookiepedia flagged as a tracking cookie. This cookie was not detected because it has a lifetime of only 30 minutes, placing it outside our longevity criterion. This confirms Cookiepedia's unsuitability for this context.
The second approach is to leverage the ML-based (tracking) cookie classifier proposed by CookieGraph~\cite{cookiegraph}, which uses a supervised classifier trained on labeled cookie data.
While methodologically more principled, CookieGraph presents a critical reproducibility limitation. 
However, the published artifact ~\cite{cookiegraph.artifact} does not include a trained model, and the training data is no longer publicly available. We contacted the authors, who confirmed that the system cannot be reproduced in its original form.

As a third option, we evaluated a rule-based heuristic used in prior web-tracking literature to identify third-party tracking cookies~\cite{Acar.tracking.2014,Koop2020InDepthEO,Urban.WWW.2020,englehardt2016online}.
We adopt these definitions to enable a basic comparison between our work and previous studies that focused on third-party tracking. 
This is justified because popular third-party cookies have migrated into the first-party context (\eg Google Tag Manager~\cite{Google2025GTMDocs}) and, thus, the heuristic might be a valid choice.
The criteria are:

\begin{enumerate}[leftmargin=*]
    \item \emph{Longevity}: The cookie is not a session cookie and has a lifetime exceeding 90 days.
    \item \emph{Length}: It contains at least eight bytes to ensure sufficient entropy to hold a user ID.
    \item \emph{Uniqueness}: It is unique in each measurement run, with the length of each value differing by no more than 25\%.
    \item \emph{Similarity}: The cookie values are similar according to the Ratcliff/Obershelp string comparison algorithm~\cite{ratcliff1988}, with a similarity threshold of 60\% or less.
\end{enumerate}

In related work, the heuristic is applied to third-party tracking cookies. In our approach, we use the heuristic on first-party tracking cookies. As discussed, first-party cookies are used for various means where an ID is required, but user tracking cannot be assumed (esp., session identification). To validate that the heuristic can be applied in this context, we (1) evaluate manually 100 identified FPT cookies (the top 50 FPT cookies and 25 randomly sampled FPT cookies to test if the heuristic correctly identifies FPT cookies; and 25 randomly sampled non-FPT cookies to test for false negatives), (2) manually analyze 15 well-known FPT cookies to determine if they are correctly classified, and (3) cross-compare the identified cookies with the CookieGraph dataset~\cite{cookiegraph, cookiegraph.artifact} to assess coverage.
For the manual analysis, we used several online sources (\eg provider documentation, open databases, and site owner) and Internet search engines (\eg by searching for the cookie name) to identify its purpose. While being time-intensive, we could find a purpose for all manually analyzed cookies.

The manual evaluation of the heuristic showed that it \emph{can} be used for identifying first-party tracking cookies.
For the 10 well-known tracking cookies, the heuristic identified one false negative and two edge cases (possible false positives). In one case (\texttt{PHPSESSID}), it appears a site may have abused PHP session IDs for user tracking.
In the other case (\texttt{AWSALB}), load-balancing mechanisms were configured with very long-lasting routing information. In both cases, cookies were configured with very long lifetimes ($>$6 months) and hold a user ID.
Further, the analysis of the 100 sampled cookies shows that 60\% are directly related to advertising, 33\% can be used for cross-site tracking but also serve other purposes (\eg bot protection), and 7\% are not directly tied to tracking. 
More detailed results of these validation steps are provided in \cref{app:heuristicstests}.
Finally, we cross-compare our heuristic classification against the publicly available CookieGraph dataset~\cite{cookiegraph,cookiegraph.artifact} to assess overlaps of both mechanisms.
This comparison yields an 70\% overlap of identified cookies.
Of the cookies identified by CookieGraph but not by the used heuristic, manual validation of prominent examples shows that at least 41\% are related to session management (\eg \texttt{\_hjSession\_}) or are valid for only one session, and are therefore out of scope for our study. We therefore conclude that reproducing CookieGraph's approach would not be expedient for our study. %

\fakeparagraph{Identifying Known Trackers}
To determine the URLs that are (knowingly) associated with user tracking, we used two popular ad-blocking and anti-tracking filter lists, namely EasyList (Version: 202505191305) and EasyPrivacyList (Version: 202505191305) \cite{EasyList.2023,EasyListPrivacy.2024}. 
We use these lists as a baseline to: (1) evaluate the efficacy of current rule-based defenses against first-party tracking, and (2) benchmark our own detection mechanisms against known tracking entities.
Filter lists identify trackers by matching URLs against a set of rules (\eg domain names), a mechanism that does not necessarily transfer to the first-party context~\cite{boettger.2025, fouad2020}. %

\subsection{Measurement Dataset Overview}
\label{sec:measurement_overview}
In our experiment, we visited 758,960 pages, collected over 477,231 first-party cookies, and stored over 3\,TB of data.
\cref{tab:general_overview} provides an overview of the measured data. 
Note that our goal with the four measurements is not to identify how different regulations (\eg GDPR) impact first-party tracking, but rather to obtain a representative dataset~\cite{demir_2023,Demir_rep_2022}.

\begin{table}[tb]
    \centering
    \resizebox{\columnwidth}{!}{
    \centering
    \begin{tabular}{lrrrrr}
    \toprule
    {\bfseries ID}
    & {\bfseries Sites}
    & {\bfseries Pages}
    & {\bfseries HTTP Req.}
    & {\bfseries Cookies}
    & {\bfseries JavaScripts}     \\
    \midrule
     EU1 & 24,976 & 192,470 & 15,822,039 & 5,017,332 & 4,200,529  \\
     EU2 & 24,975 & 192,322 & 15,742,516 & 4,985,296 &  4,175,046 \\
     US1 & 23,533 & 181,689 & 21,792,272 & 5,966,304  & 9,831,813 \\
     US2 & 24,981 & 192,479 & 22,936,063 & 6,490,947 & 10,160,264 \\
    
    \bottomrule
    \end{tabular}

     }
     \caption{Overview of the measurement dataset.}%
     \label{tab:general_overview}
 \end{table}

\fakeparagraph{Observed Cookies}
Overall, we collected 309,887 distinct cookie names and 577,963 distinct cookies (name, path, origin), of which 477,231 were first-party and 151,395 third-party. We identified 50,665 cookies that appear in both contexts across profiles; since each cookie is treated individually, this overlap has no impact on our analysis.
On average, each profile contains 5,614,969 (min:~4,985,296, max:~6,490,947, SD:~740,367) cookies of which 1,060,511 were classified as first-party cookies, we exclude all third-party cookies from further analysis.
On average, each page includes 36 (min:~1, max:~11,319, SD:~92) cookies. 466,740 pages contains more than 10 cookies.

\fakeparagraph{Identifying Potential First-Party Tracking Cookies}
As described in \cref{sec:measurement_overview}, we found ~470k first-party cookies, and applied the defined heuristic (see \cref{sec:cookies_identify}) to identify potential first-party tracking cookies.
After removing cookies with a too short lifespan and a too short length (criteria 1 longevity and 2 length), we have 452,380 distinct cookies and 33,072 distinct cookie keys.
Next, we verify whether a cookie appears in more than one of our measurement profiles, so we can test criteria 3 (uniqueness) and 4 (similarity). 19,533 (59\%) cookies were filtered out in this step.
In the next step, we check if a cookie value is unique and has a similar length across all profiles (criterion 3 \emph{uniqueness}).
We removed 1,432 (4.3\%) of the remaining cookies during this step. 
Finally, we check the similarity of the cookies across the profiles (criterion 4 similarity) and removed 9,712 (29.3\%) cookies in this step.
After applying the heuristic, we identified 6,249 distinct cookie names that potentially hold an ID.
In our final dataset, we included all cookie keys if they were once classified as a potential tracking cookie (\ie holding an ID). %
This step ensures that if a cookie of interest is set with a placeholder value (\eg an empty string or dummy value), it is not excluded due to criteria 2, 3, or 4 of the heuristic. 
Our data showed that such behavior is not rare, and omitting this step would eliminate cookies such as \texttt{\_ga} (Google tracking cookie) and \texttt{\_fbp} (Facebook tracking cookie).  

To homogenize the analyzed cookie names, we use known cookie name patterns provided by \nolinkurl{cookies.is}~\cite{cookie.isDatabase} (\eg \texttt{\_ga\_<USERID>} or \texttt{AMCV\_<ORGID>@AdobeOrg}), to unify the cookie names (\eg changing \texttt{\_ga\_<USERID>} to \texttt{\_ga}).
This step allows us to provide a better, more interpretable overview of first-party cookie use, as cookie names are clustered.
Afterward, we have a total of 25,136 (76\%) first-party cookie names that might be used for FPT. %

\fakeparagraph{Collected JavaScript Code}
To analyze the server-side tracking ecosystem, we collected the JavaScript code, the parties that delivered it, and the contexts in which it was included.
Overall, we stored 6,280,920 JavaScript files with 2,431,533 distinct MD5 hashes delivered by 24,911 parties (14,535 first parties and 10,376 third parties). 
Of these scripts, 639,947 interacted with a cookie (\eg accessing or setting it) that we identified as a first-party tracking cookie (EU1:~337,884, EU2:~288,005, US1:~144,535, US2:~181,385).
We exclude all other scripts from our analysis.

To identify which party interacts with a first-party tracking cookie, we classify each script by a tuple $(d, s, t)$ of loading page, delivering party, and top-level site. A script is first-party if the eTLD+1 of $s$ matches $t$; a cookie is set in a first-party context if $d$ equals $t$, regardless of whether $s$ is a third party. In FPT deployments, a first-party may directly include a third-party tracking script (see \cref{sec:background}), making cross-site script similarity a relevant indicator of shared tracking infrastructure.

To understand whether similar tracking scripts are reused across different sites, we assess the structural similarity of all collected scripts. Identifying such reuse is essential because, in the first-party context, the hosting domain alone cannot determine whether a script is tracking-related; content-based analysis is required (see~\cref{sec:overview_sst}).
We use SimHashes and the SimHashIndex~\cite{SinghJain2007} to assess the similarity of all scripts collected for our analysis. Based on their SimHash, we found 291,542 %
distinct file contents over all profiles.

\section{The First-Party Tracking Ecosystem}
\label{sec:sst_ecosystem}
In this section, we examine the FTP ecosystem by identifying potential FPT tracking scripts, clustering them by source code, attributing them to entities, and assessing the ecosystem's structure.

\subsection{Overview of Identified JavaScript}
\label{sec:overview_sst}
First, we provide an overview of the scripts used for FPT purposes to highlight their use in the field and assess whether migration patterns from third-party to first-party contexts are observable.
The similar script code might be hosted and delivered directly by various first parties (see \cref{sec:background}) or a third party. 
Thus, to identify the tracking code, one cannot simply rely on the party delivering the script; one must assess its content. 
To establish an understanding of the structural similarities and uniqueness of the collected scripts, we compute SimHash~\cite{charikar2002simhash,SinghJain2007} fingerprints for them. SimHashes allow us to compare similarities between scripts (\eg scripts that only differ in a unique identifier that specifies the hosting party will have a similar or the same SimHash. %
This comparison enables us to quantify the degree of similarity and divergence between the observed scripts across all sites in the measurement profiles (see \cref{sec:cluster_sst}). 
We choose SimHash over abstract syntax trees (AST), as AST-based approaches incur $\mathcal{O}(n^3)$ overhead versus $\mathcal{O}(n)$ for SimHash, with no benefit for our use case; SimHash has further been established for source code comparison~\cite{WebAlmanac.2024.Privacy}.

First, we compare the number of distinct SimHashes of each profile. 
Note that we excluded all JavaScript resources that do not interact with the identified potential first-party tracking cookies (see \cref{sec:measurement_overview}).
Overall, we found 26,606 distinct scripts, according to their SimHash. 
Of those 12,073 (45\%) were delivered in the first-party context and 15,451 (58\%) in the third-party context.
The numbers do not add up to the total number of identified scripts because we found 916 (3.5\%) SimHash-identical scripts that are hosted in both first-party and third-party contexts, indicating that in some cases, the first party directly delivers a script that interacts with potential first-party tracking cookies, which are also delivered by a third party.
Overall, these numbers indicate that potential FPT-tracking scripts are hosted in both contexts, suggesting that tracking is no longer limited to third parties. Furthermore, these numbers suggest that trackers are increasingly moving into first-party contexts, making blocking more challenging.

\begin{figure}[tb]
    \centering 
    \includegraphics[width=\columnwidth]{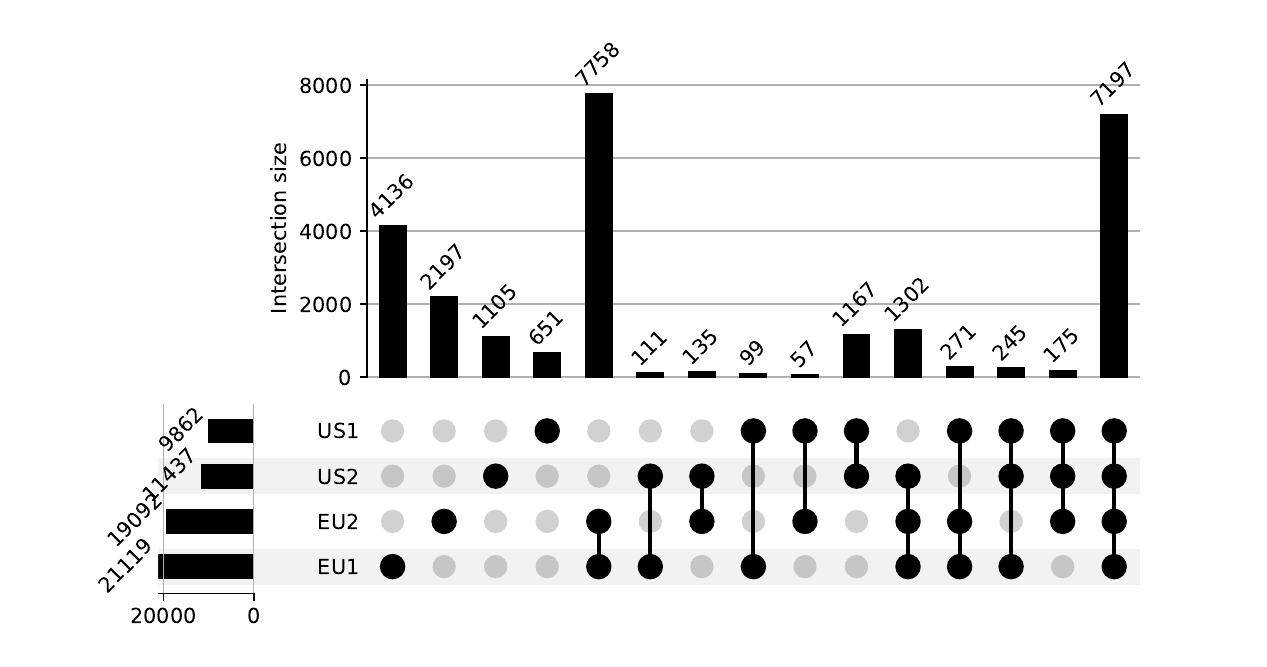}
     \caption{Exclusive intersections between the observed scripts in the profiles. The bars on the lower-left show the total size of each dataset. The top bars, in combination with the point matrix, show the intersection sizes for each profile. } %
    \label{fig:upset_plot}
\end{figure}

\Cref{fig:upset_plot} provides an overview of the UpSet analysis~\cite{lex2014upset} of SimHashes in each profile. 
The UpSet analysis visualizes shared and unique elements across multiple datasets by quantifying their intersections. In contrast to Venn diagrams, it scales efficiently to a larger number of sets and highlights both overlapping and exclusive relationships. The UpSet analysis demonstrates both strong overlaps and distinct differences between the measured datasets. 
Both regional pairs show substantial overlap (EU: 16,528 shared scripts; US: 8,784), with cross-region intersections of comparable magnitude (7,7k to 7,8k).

The results highlight a core of shared scripts across both EU and U.S. datasets, suggesting that tracking mechanisms are globally deployed. Further, the asymmetric differences indicate regional or context-dependent scripts (\eg due to JavaScript bundling), potentially reflecting variations in website audiences, regulatory environments, or implementation details of tracking methods.

\subsection{Clustering FPT JavaScript Code}
\label{sec:cluster_sst}
As discussed, the same or highly similar code is hosted by first- or third-party providers. 
Thus, to assess whether tracking code is reused, the URL cannot serve as an indicator to identify FPT-related requests.
To gain a deeper understanding of FPT and the entities involved, we cluster the JavaScript code that interacts with potential FPT cookies.

\fakeparagraph{Method to Cluster JavaScript Code}
Based on the computed SimHashes of the identified FPT-related JavaScripts, we build the SimHashIndex~\cite{SinghJain2007}, which clusters scripts based on their source-code similarities. We modified the SimHashIndex Python library~\cite{1e0ng_simhash} to store the SimHashes in combination with a 3-tuple: (1) script URL from which the script was loaded, (2) the top-level URL where the script is included, and (3) whether the script is a first-party script. Our SimHashes have a fingerprint bit length of $f=64$ bits~\cite{SinghJain2007,Sood.11}. To determine script similarity, we use the Hamming distance~\cite{hamming1950} with $k=8$, meaning two scripts are considered similar if their SimHash fingerprints differ by eight or fewer bits~\cite{SharifRox2011,Sood.11,SinghJain2007}, following the near-duplicate detection approach for Web crawling~\cite{SinghJain2007,Sood.11}.

Clustering operates only on distinct SimHash values; duplicate scripts are assigned to their matching cluster post-hoc. Each cluster is then linked to its site data, allowing a cluster to span multiple sites and delivery parties.

\fakeparagraph{Overview of Clustered JavaScript Code}
Overall, we found 24,914 cluster with a mean size  of 68 (min:~1, max:~186,634, SD:~1,812, median:~4) site data (\ie sites on which scripts from a cluster were loaded).
It is notable that the cluster size is typically 1 or 2 (93\% of all clusters), indicating that most clusters contain only a few distinct JavaScript files used by multiple sites. From those clusters, 95 (0.35\%) use bundling, meaning a bundle of multiple JavaScripts. %
Higher $k$ values could lead to more scripts per cluster; however, as noted in prior work~\cite{SinghJain2007,Sood.11}, this would result in scripts that are no longer similar to one another. The number of scripts in the cluster provides an initial insight into the FPT tracking ecosystem, where some tracking scripts are used across multiple sites. While others are unique on distinct sites.
It should be noted that the goal in this process is not to build large clusters, but to identify code re-usage (\ie we do not aim to dissect JS-bundles~\cite{Rack2023}).

\begin{figure}[tb]
    \centering 
    \includegraphics[width=.8\columnwidth]{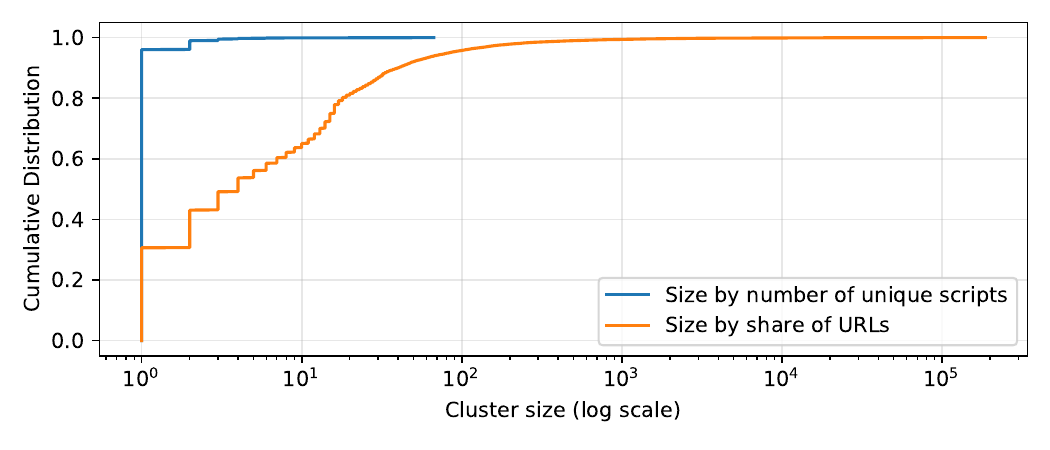}
      \caption{Distributions of the number of unique scripts (\ie SimHashes) in each cluster (blue line), and the number of URLs per cluster (orange line).}
    \label{fig:cluster_cdf}
\end{figure}

\Cref{fig:cluster_cdf} provides the distribution of clusters by their sizes, measured by URLs in a cluster. The CDF chart indicates that the vast majority of clusters are relatively small, meaning the scripts are typically confined to a single page. In the context of FPT, some scripts are widely used across multiple sites. This indicates some dominant players in the FPT ecosystem.
Specifically, 31\% of the clusters contain only a single element of site data. Meaning that unique scripts (in terms of their SimHash) are used in these clusters. From those clusters, only 24 (0.09\%) uses bundling. %
The numerous small clusters represent site-specific or rarely reused scripts, while the large clusters correspond to widely deployed, nearly identical scripts. %
The combination of a small median (4) and a large standard deviation (1,812) relative to the mean (68) indicates a right-skewed, heavy-tailed size distribution ($\mathrm{CV} \approx 26$), 
implying that a few mega-clusters~\cite{englehardt2016online} exist.
This corresponds to observations that the third-party tracking ecosystem is dominated by a few central players~\cite{Binns2018,falahrastegar2014anatomythirdpartywebtracking,Falahrastegar2017,Razaghpanah2018}.

\subsection{Attributing FPT Clusters}
\label{sec:analyzing_sst}
FPT decreases transparency as it is not easy for users to understand who (\eg which third party) is serving tracking scripts (see \cref{sec:background}).
In the following, we aim to attribute the identified clusters and their corresponding tracking scripts to third parties (\ie known trackers), even if they were served in a first-party context.
The created clusters contain scripts loaded in the first- and third-party contexts, allowing us to determine which third party is responsible for distributing a script.
Since third-party script URLs directly reveal the hosting domain, and thus the responsible provider. We use them to identify which companies operate the scripts contained in each cluster.
To identify the tracking provider, we use the \emph{WhoTracksMe}~\cite{whotracksme} database. Thus, we attribute a cluster to a third party based on the third-party domains from which the script is loaded. WhoTracksMe is based on the open-source project trackerdb~\cite{ghostery_trackerdb}, which can be incomplete. Therefore, we may not be able to attribute all providers.

If the third-party URL is related to a provider (\eg Google), we can attribute a cluster to one provider. If there is more than one provider, we assign the cluster to the provider with the largest share of script URLs, which was the case for 127 (0.5\%) clusters.
We identify 10,399 (42\%) clusters that do not contain any third-party script. Among the clusters that include at least one third-party script, our attribution method successfully assigns 8,477 (34\%) clusters to a provider. Applying the same approach on the first-party-only clusters yields additional 663 (3\%) attributed clusters. Overall, this results in 9,140 (37\%) clusters that can be attributed to a specific provider.
Although the WhoTracksMe database does not map every observed eTLD+1 to a third party, our attribution method still classifies 37\% of all clusters as belonging to a tracking entity. These attributed clusters encompass 76\% (91,209) of all measured script URLS. The results indicate that a substantial share of tracking infrastructure operates without visible third-party scripts, highlighting how FPT reduces transparency by shifting functionality to first-party code. Further, it shows that large providers dominate the ecosystem, while many smaller on-server-side implementations remain difficult to detect.

\fakeparagraph{Top Cluster in the FPT Ecosystem}
Due to the heavy-tailed distribution of entity sizes within the tracking ecosystem, a small number of actors account for a disproportionate share of the overall activity. The top 10 clusters represent 44\% of the script URLs and are present on 74\% of the sites and 72\% pages. Examining the top 10 clusters based on site data thus provides a representative view of the dominant entities that shape the ecosystem's structure and dynamics, while maintaining analytical tractability.
For these clusters, 99\% of the scripts are also loaded in a third-party context (see \cref{tab:top_cluster}), allowing provider attribution via third-party URLs. %

In \cref{tab:top_cluster}, we provide an overview of the analysis and attribution for the top 10 clusters. In total, we found 52,822 distinct URLs in the top 10 clusters. 
Google is the prominent actor in these clusters, which aligns with previous observations of the Web tracking ecosystem~\cite{englehardt2016online, bonfils2025empiricalinquirysurveillancecapitalism}.
Aggregated over all clusters, 8\% of the requests are issued by the first-party context, showing that a substantial part of tracking is already happening in this context. 
In the top 10, only 2 clusters (overall 13,534; 50\%) operate solely in the third-party context, highlighting a trend toward FPT.
In these cases (\ie Meta and HubSpot), the tracking scripts are loaded by a third-party request, but the scripts utilize FPT cookies.
Further, we see no direct relation between the number of sites that employ a specific script from a cluster and the share of first-party requests. This observation suggests that there is not a small share of clusters responsible for most FPT, but rather several clusters. %
However, the ecosystem seems to be dominated by central players.

\begin{table*}[tb]
    \centering
    \resizebox{\textwidth}{!}{
        \centering
\begin{tabular}{clrrrrrrr}
\toprule
{\bfseries No.} &
{\bfseries Attribution} &
{\bfseries \# Sites} &
{\bfseries \# Pages} &
{\bfseries Cookies Used} &
{\bfseries Script URLs} &
{\bfseries Requests} &
{\bfseries 1st Party Req.} &
{\bfseries 3rd Party Req.} \\
\midrule
1  & Google & 5{,}141 & 44{,}954 & 597 & 14{,}611 & 186{,}634 & 1{,}411 & 185{,}223   \\
2  & Google & 2{,}990 & 27{,}916 & 489 & 7{,}599  & 125{,}637  & 1{,}131 & 124{,}506  \\
3  & Google & 4{,}311 & 36{,}467 & 523 & 33       & 81{,}832  & 292     & 81{,}540    \\
4  & Google & 2{,}889 & 22{,}379 & 521 & 11{,}106 & 81{,}094   & 234     & 80{,}860   \\
5  & Google & 3{,}532 & 28{,}270 & 614 & 9{,}162  & 75{,}940   & 1{,}198 & 74{,}742   \\
6  & Google & 2{,}121 & 18{,}013 & 389 & 5{,}461  & 55{,}799   & 77      & 55{,}722   \\
7  & Meta & 3{,}458 & 31{,}955 & 518 & 7 & 52{,}186  & 0       & 52{,}186   \\
8  & Google & 1{,}252 & 10{,}688 & 283 & 2{,}894  & 44{,}395   & 572     & 43{,}823   \\
9  & HubSpot & 555& 5{,}077 & 123 & 1{,}785  & 24{,}356    & 0       & 24{,}356   \\
10 & Google & 678     & 6{,}585 & 150 & 1{,}177  & 20{,}485    & 334     & 20{,}151  \\
\midrule
\multicolumn{2}{l}{Across \emph{all} clusters} & 13,187 & 136,891 & 3,378 & 119,912 & 1,705,780 & 142,036 & 1,563,695 \\

\bottomrule
\end{tabular}

    }
    \caption{Top 10 clusters by the number of observed HTTP requests loading potential FPT-related JavaScripts.}
    \label{tab:top_cluster}
\end{table*}

\fakeparagraph{Attribution of First-Party Provider}
Overall, we could attribute clusters to 492 different providers. \cref{tab:top_attr} provides an overview of the top 10 providers, which covers 5,448 (60\%) of the attributed clusters. It is notable that Logicad (a demand-side platform used for real-time bidding) is the most attributed cluster, but overall, clusters attributed to companies such as Google, Adobe, or Amazon contain more sites, thereby increasing their relevance to the FPT ecosystem.
Among the top 10 providers, Google has high coverage across sites and pages where the clusters are present, indicating its reach in the FPT ecosystem.
The results show that tracking providers known from the third-party context are also dominant in first-party tracking. 
Overall, the number of clusters per entity exhibits a long tail distribution.
While these partners contribute fewer clusters overall (roughly 14\% of all clusters), they are included on a substantial number of sites (\eg Microsoft with 2,717 sites).

\begin{table}[tb]
    \centering
    \resizebox{\columnwidth}{!}{
        \centering
\begin{tabular}{clrrr}
\toprule

{\bfseries No.} &
{\bfseries Provider} &
{\bfseries Clusters} &
{\bfseries Share} &
{\bfseries Sites} 
 \\
\midrule

1  & Logicad &	1,675 & 30.75\% & 35\\
2  & Google	 & 1,555 & 28.54\% & 10,432\\
3  & Adobe   &	567 & 10.40\% &1,107 \\
4  & Wingify &	461 & 8.46\% &175\\
5  & Amazon Associates &	442& 8.11\% &1,246\\
6  & Tealium & 295 & 5.41\% &258\\
7  & Alibaba & 133 & 2.44\% &182\\
8  & HubSpot & 128 & 2.39\% &730\\
9  & Siteimprove &	98& 1.80\% &149\\
10 & Microsoft	& 94& 1.73\% &2,717\\

\bottomrule
\end{tabular}

    }
    \caption{Top 10 of the actors to which we found the most attributed clusters, ordered by cluster size.}
    \label{tab:top_attr}
\end{table}

\subsection{Usage of First-Party Tracking Cookies}
We examine the most commonly used cookies and the distribution of cookies from first- and third-party scripts across the different clusters.
To establish a consistent basis for comparing cookie usage across clusters, we first normalize cookie names.
In many cases, cookie providers augment the base cookie name with suffixes (\eg \texttt{\_ga\_<ID>}), which obscures their semantic equivalence. We therefore extract the shared prefix of each cookie by removing these identifiers, \eg transforming \texttt{\_ga\_<ID>} into \texttt{\_ga}. Following related work, we employ a pattern-list and wildcard-based approach to derive normalization rules~\cite{Xuehui2021}. We extracted the patterns from \nolinkurl{cookie.is}~\cite{cookie.isDatabase} and applied them to our database to identify and collapse cookies into their canonical form. Using this method, we successfully normalized 37,647 cookies. 

\fakeparagraph{Cookie Usage by Clusters}
After normalization, we examine the distribution of cookie usage across the identified clusters. On average, a cluster uses 3 (min:~1, max:~614, SD:~13, median:~1) first-party tracking cookies, and 11,071 (44\%) of the clusters use more than one cookie. 
56\% of clusters use only a single cookie. These single-cookie clusters predominantly rely on \texttt{\_\_smn\_fid} (6.7\%), \texttt{\_gcl\_au} (5.6\%) from Google, \texttt{\_abck} (4.7\%) from Akamai, or \texttt{\_ga} (1.5\%) also from Google. In total, we identify 1,525 distinct cookies that occur exclusively in a single cluster. 13,749 (55\%) of such clusters contain a single script responsible for setting that cookie, indicating highly customized FPT and SST deployments. The \texttt{\_\_smn\_fid} cookie, used by Search Marketing Networks (SMN) (\eg Google Ads, Microsoft Advertising), exemplifies such tailored infrastructures. Notably, clusters with only one cookie have up to 5 unique JavaScripts that set it. Furthermore, 6,595 (27\%) of these clusters rely solely on first-party scripts, meaning that roughly one-quarter are first-party–only configurations in which multiple unique scripts set a single identifier. Thus, this practice obscures detectability and decreases transparency for users. %
Additionally, 7,192 (29\%) clusters without first-party scripts, indicating continued third-party dominance. Finally, only 56 (0.2\%) of scripts in these clusters were loaded within a first- \emph{and} third-party context, cases where a migration from third- to first-party could be happening. These clusters include 123 sites and 745 pages. Totally, clusters with only one cookie include 5,789 (24\%) sites and 35,087 (5\%) pages from our scope, where first-party only clusters have one percent more sites and pages than third-party only clusters. In FPT, a small set of cookies dominate clusters, including just one cookie. Most of them are attributed to a well-known third-party provider. This suggests that third-party tracking providers also play a significant role in the first-party tracking ecosystem.

Among clusters using more than one cookie, 3,804 (15\%) load all cookies exclusively in the first-party context, whereas 6,342 (25\%) load them only in the third-party context. Third-party scripts occur 4 times more frequently, yet third-party code still influences tracking workflows even in first-party contexts. 
In the long tail, third-party context may  slightly shift out, as first-party becomes more dominant, decreasing future third-party script influence. 
Further, we find identical scripts and cookies in both contexts illustrate the intertwined FPT mechanisms. 925 (3.7\%) of the clusters that include more than one cookie are loaded in the first- \emph{and} third-party context. The ratio of third- to first-party URLs in those clusters is $23:1$, indicating strong dominance of third-party URLs. Clusters that have cookies loaded in first- and third-party include 12,003 (49\%) sites and 123,102 (16\%) pages, indicating migration is significant. The number of tracking cookies across multiple clusters on nearly half of the sites indicates that third-party tracking providers dominate FPT. 

\fakeparagraph{Cookie Cross Usage}
Finally, we examine relations between cookies by studying their co-occurrence.%
The most common tuple combines the two most frequently used cookies overall. However, the intersection, only 11.5\% of clusters containing both, reveals considerable functional heterogeneity in FPT deployments. Despite the broad adoption of individual cookies, their limited coexistence suggests that provider-specific, rather than universal, configuration strategies are used. This results in substantial configurational diversity, where specific combinations of identifiers align with distinct tracking workflows. Across sites, the top ten pairs appear in 47\%–50\% of all domains in our measurement scope, highlighting the widespread operational relevance of cookie pairings in the FPT ecosystem.
These cross-cookie usage results map to and reduce findings of Bahrami \etAl~\cite{NikkhahBahrami2025}.

\subsection{Structure of the FPT Ecosystem}
\label{sec:sst_graph}
To characterize the relational dependencies within the FPT ecosystem beyond scripts and cookies, we construct a network graph from the clustered JavaScript and observed cookies to reveal the emergent topology of FPT deployments.
We define a node as a cookie or JavaScript cluster. Edges between cookies and clusters are built if a cookie is used  (\ie set or accessed) by a script in a cluster.
Appendix~\ref{app:ecosystem} provides an overview of the top 5 clusters by site data.

The graph contains 28,292 nodes with 72,097 edges and has an average connectivity of 2.5 edges per node. 
Overall, the graph is sparse, with only a small fraction of all possible cluster-cookie relationships. This observation indicates selective cookie reuse rather than arbitrary identifier deployment. From an ecosystem perspective, this structure implies that FPT is highly asymmetric, focusing on a small subset of cookies, and clusters likely enable disproportionate cross-site linkage. The ecosystem itself contains 162 components with the largest component of 25,901 nodes. 
This implies that clusters and cookies are structurally connected through shared vendors or infrastructure.
From the perspective of privacy developers and researchers, this finding is a first indication that novel defense mechanisms, such as cookie or JavaScript-based ones, may be a path towards protecting users. We leave this for future work. 
Isolated components represent custom, first-party-only deployments with no dependencies. In FPT, clusters embedded within large components pose a higher privacy risk due to indirect identifier linkage. 

To further analyze the structural dependencies between identifiers, we project the cluster-cookie graph into a one-mode cookie-cookie graph, where edges denote the co-occurrence of two cookies within the same cluster. While the original graph is highly sparse, the resulting projection exhibits a pronounced edge explosion, yielding 3,181 cookie nodes connected by 625,486 edges. This indicates that cookies are not deployed independently but rather as tightly coupled bundles, where even moderately sized clusters induce near-clique structures in the projection. Applying modularity-based community detection to this graph reveals that detected communities primarily correspond to recurring cookie bundles rather than isolated identifiers. These communities capture standardized FPT configurations that are reused across multiple clusters and sites, reflecting shared infrastructure or template deployments. From a privacy perspective, this shows that tracking capability emerges from the collective behavior of co-deployed cookies, suggesting that reasoning about individual identifiers in isolation underestimates the effective tracking surface exposed by FPT systems.

\section{Protecting Against FPT}
\label{sec:sst_blocking}
As shown, tracking activities are increasingly moving into the first-party context. 
This opens the question of whether current rule-based approaches sufficiently protect users' privacy.
To advance FPT detection, we propose a data-driven rule-based classification technique that derives blocking rules from structural patterns in URLs. 
Our approach uses patterns in URL query parameters, which are often indicative of tracking-related activity. 
We evaluate the rules and test if they can detect FPT-related URLs without breaking pages.

\subsection{Dataset Construction}
\label{sec:dataset_construction}
First, we assess whether current blocking lists are sufficient to protect against FPT.
Therefore, we test how many of the identified URLs that serve FPT-related JavaScript (see \Cref{tab:top_cluster}) are flagged by EasyList or EasyPrivacy. 
We find that only 33,081 (27,5\%) of the URLs would have been blocked by any of these two lists.
Thus, current protection mechanisms do not yet provide adequate coverage for these emerging tracking technologies. 
To overcome this limitation of the blocklists, we develop an approach to derive rules tailored for FPT based on URL features.
To build protection mechanisms against FPT, we curate two datasets ensuring that mined patterns reflect FPT-specific behavior rather than incidental measurement properties.

Our analysis relies on two datasets collected during the large-scale measurement. 
The first dataset contains FPT-related URLs that served JavaScripts that are \emph{not} flagged by either list.
We chose our approach so that it captures criteria that are not yet covered and does not identify patterns already present in current lists.%
The dataset contains 86,831 URLs. 
The second dataset contains the same number of randomly sampled ULRs that are \emph{not} related to tracking, as determined by both our approach and the two lists.  
For both datasets, we used 90\% of the URLs for training (78,147 URLs per dataset) and 10\% for validation (8,684 URLs).
We name the datasets that contain the FPT-related requests $\mathcal{D}_{FPT}$ (training) and $\mathcal{V}_{FPT}$ (validation), and accordingly the datasets that contain the non-tracking-related requests $\mathcal{D}_{Other}$ and $\mathcal{V}_{Other}$.

\subsection{Feature Extraction}
Naturally, in FPT, the domain name in a URL is a weak indicator for tracking, as it changes from site to site.
Therefore, to test whether FPT-related requests exhibit consistent structural patterns at the URL level, we only consider the set of query-parameter keys for our analysis (\eg parameter names such as \texttt{cid}). 
Concretely, we extract all query keys from $\mathcal{D}_{FPT}$ and treat each key as a binary feature indicating its presence in a request.
We observe 940 unique query keys, forming a highly skewed vocabulary. %

\begin{figure}[tb]
    \centering
    \includegraphics[width=.8\columnwidth]{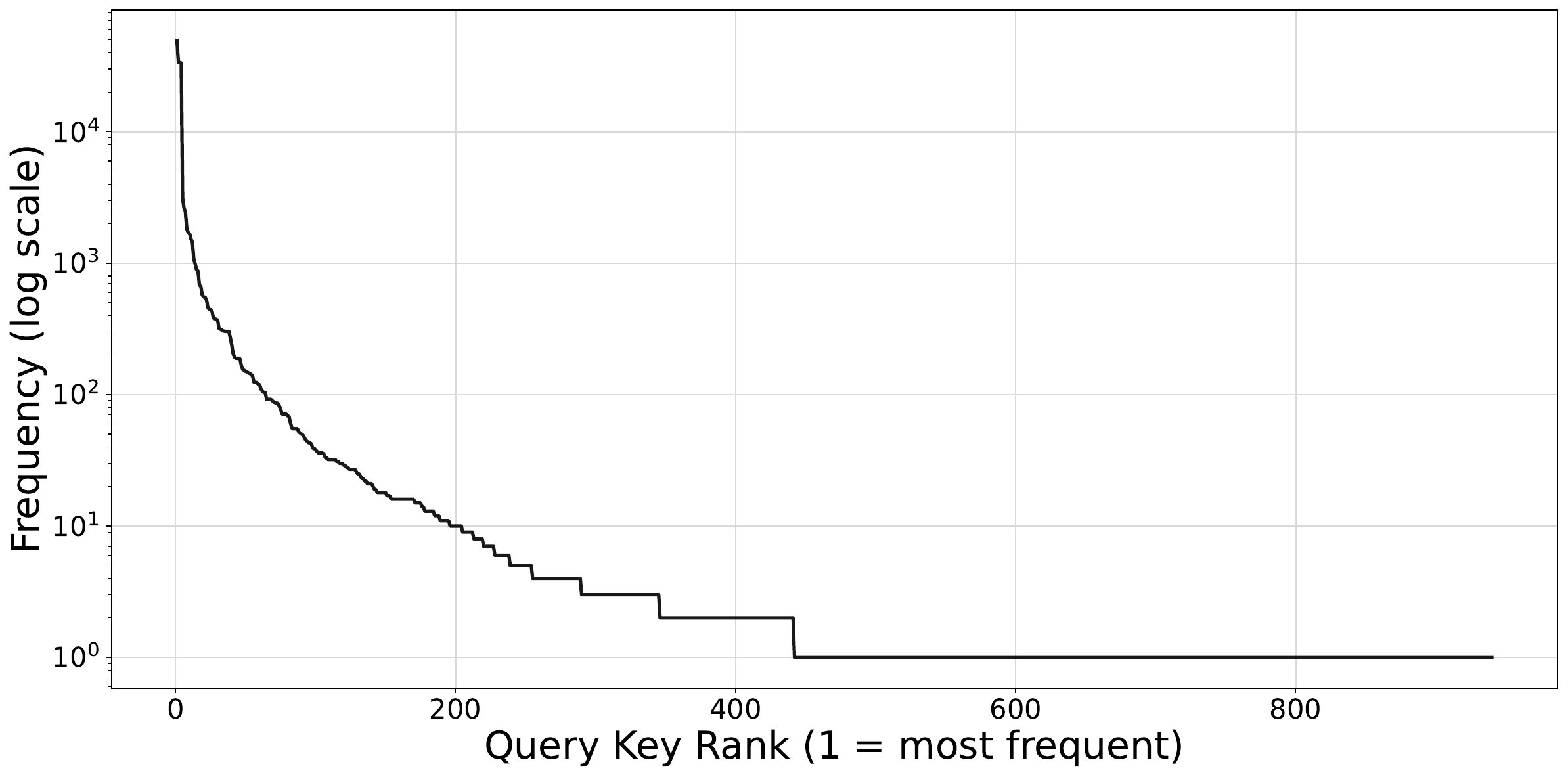}
    \caption{Rank–frequency distribution: keys are ordered by decreasing frequency. Note the logarithmic scale.}
    \label{fig:Rank–Frequency}
\end{figure}

\begin{figure}[tb]
    \centering
    \includegraphics[width=.8\columnwidth]{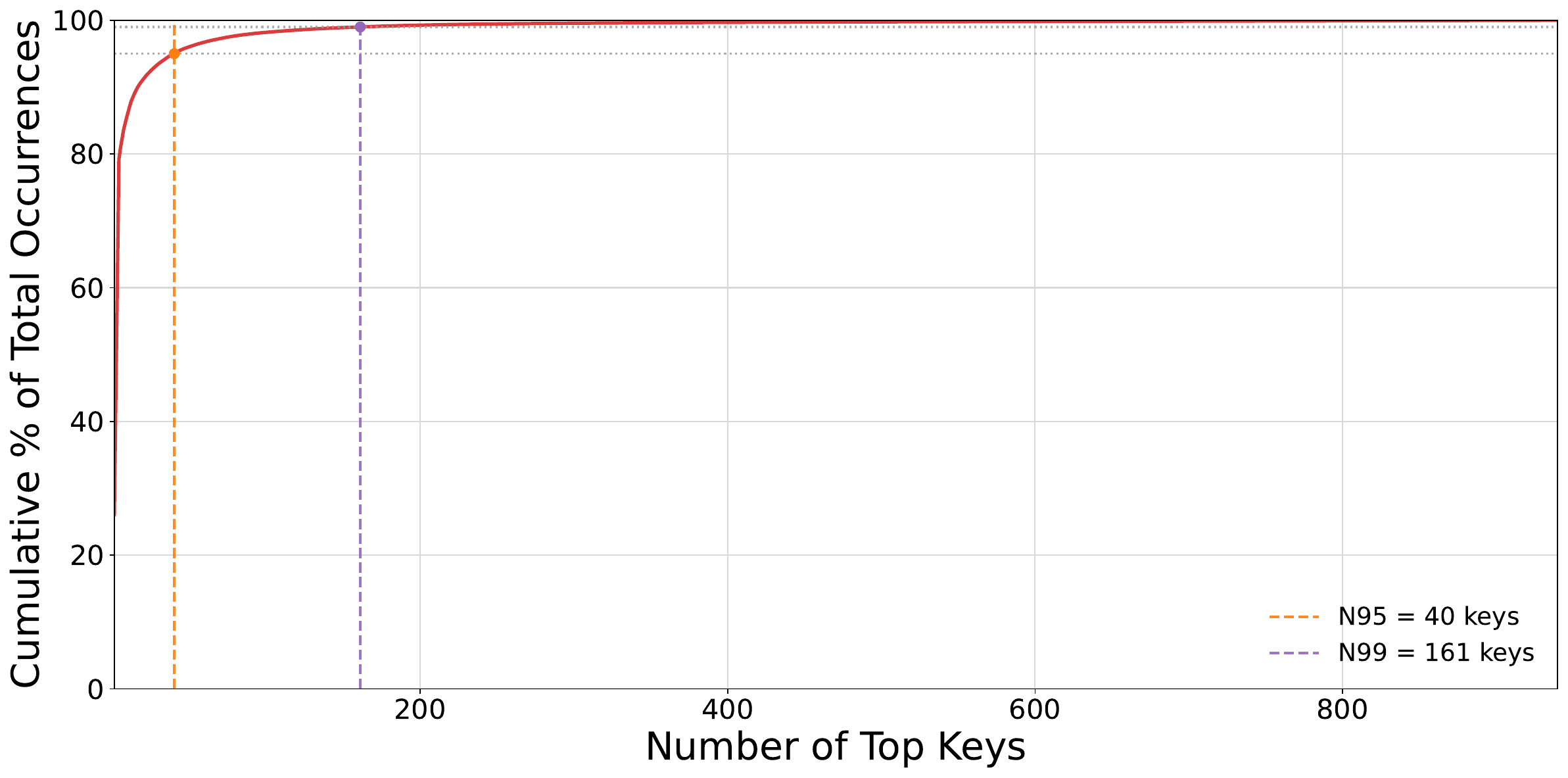}
    \caption{Cumulative coverage curve for the identity keys.} %
    \label{fig:cumulative_coverage}
\end{figure}

The rank-frequency curve (see \Cref{fig:Rank–Frequency}) exhibits a rapid decay on a logarithmic scale: a small subset of keys accounts for most occurrences, whereas a long tail of keys appears only sporadically. 
This Zipfian-like distribution suggests that most structurally informative variation is concentrated in the high-frequency region.
Complementing this, the cumulative coverage curve in \Cref{fig:cumulative_coverage} highlights the strong concentration of key occurrences: the top 161 keys account for approximately 99\% of all key appearances. %
This Pareto-like property motivates restricting the feature space to the most frequent keys as a practical trade-off between coverage and feature dimensionality.
Further, this restriction will yield fewer rules for more popular keys, bridging a trade-off between blocking lists that contain too many rules that are rarely or never used~\cite{boettger.2025}.
Based on this selection, we construct augmented datasets by adding binary indicator columns for each retained key. 
To prepare each training example (URL) for pattern mining, we tokenize its query string into query keys and retain only those keys contained in the selected vocabulary. 
Thus, each request is represented by a 161-dimensional binary feature vector indicating which keys are present.

\subsection{Frequent Pattern Mining}
Before mining feature combinations at scale, we performed an exploratory correlation-gap analysis as a proof-of-concept to verify that FPT-related traffic exhibits class-specific \emph{co-occurrence} structure among query keys. Specifically, we selected the top 100 most frequent query keys and computed their pairwise association structure (a correlation matrix) separately for tracking and non-tracking requests.
To quantify how strongly a candidate pattern $p$ separates the two classes, we define the \emph{Correlation Gap} (CorrGap) as
\begin{equation}
\text{CorrGap}(p) = \left| \Pr(p \mid \text{FPT}) - \Pr(p \mid \text{Other}) \right|
\end{equation}

Here, $\Pr(p \mid \text{tracking})$ denotes the empirical probability that a tracking-labeled request (\ie all requests in $\mathcal{D}_{FPT}$) contains pattern $p$, and $\Pr(p \mid \text{non-tracking})$ is defined analogously for requests in $\mathcal{D}_{Other}$. We estimate both probabilities as relative frequencies over the respective datasets. In this proof-of-concept analysis, $p$ corresponds to a key pair $(k_i,k_j)$, \ie the joint occurrence of two query keys within the same request. A higher CorrGap score indicates key pairs whose co-occurrence is substantially more (or less) common in FPT-related traffic than in other traffic.
Empirically, this analysis showed a clear separation between tracking and non-tracking traffic for several query-key pairs, confirming the existence of a class-specific co-occurrence structure. %

As this pairwise analysis does not scale to thousands of features and multi-million-row datasets, we transition to scalable frequent pattern mining.
We represent each HTTP request as a transaction containing the set of query keys present in its URL, and mine frequent itemsets using FP-Growth~\cite{han2000} (Frequent Pattern Growth). FP-Growth is well-suited for our setting because it avoids explicit candidate generation and remains tractable on large, sparse transactional data by compressing requests into an FP-tree and recursively mining conditional subtrees.
We mine itemsets separately on $\mathcal{D}_{FPT}$ and $\mathcal{D}_{Other}$, restricting the maximum itemset size to 3 keys to focus on short, interpretable structural interactions while limiting combinatorial growth.
Further, analyzing more keys would lead to more complex blocking rules, which might be too slow to process at scale in a state-of-the-art ad blocker.
We set the minimum support to 0.001 (\ie an itemset must occur in at least 0.1\% of requests) to retain only robust patterns that generalize beyond individual websites. Under this configuration, FP-Growth finds 456 frequent itemsets across both datasets. %

\subsection{Statistical Scoring and Rule Selection}
\label{sec:sst_scoring}
FP-Growth returns frequent itemsets in a dataset, but frequency alone does not imply that a pattern is specific to FPT.
Therefore, we apply a contrastive scoring procedure that quantifies, for each candidate itemset $I$, how strongly it separates FPT-related from non-tracking requests.
Let $\mathcal{D}_{FPT}$
and $\mathcal{D}_{Other}$
denote the tracking and non-tracking training datasets, respectively, and let $\mathbf{1}[I \subseteq r]$ indicate whether request $r$ contains all keys in itemset $I$. We compute class-specific supports as empirical probabilities:
\begin{equation}
    \begin{aligned}
        \mathrm{support}_{\mathrm{FPT}}(I)
        &= \frac{1}{| \mathcal{D}_{FPT}|} \sum_{r \in \mathcal{D}_{FPT}} \mathbf{1}[I \subseteq r], \\
        \mathrm{support}_{\mathrm{Other}}(I)
        &= \frac{1}{| \mathcal{D}_{Other}|} \sum_{r \in \mathcal{D}_{Other}} \mathbf{1}[I \subseteq r].
    \end{aligned}
\end{equation}

\noindent
We additionally record the corresponding absolute counts $c_{\mathrm{FPT}}(I)$ and $c_{\mathrm{Other}}(I)$.
To quantify enrichment in tracking traffic, we define the \emph{support ratio} as follows:
\begin{equation}
\mathrm{SR}(I)=\frac{\mathrm{support}_{\mathrm{FPT}}(I)+\epsilon}{\mathrm{support}_{\mathrm{Other}}(I)+\epsilon}
\end{equation}
where $\epsilon$ is a small constant to avoid division by zero when an itemset never occurs in the  $\mathcal{D}_{Other}$ (\ie a set of keys only occurs in tracking requests). We also report $\log_2 \mathrm{SR}(I)$ as an effect size on a multiplicative scale.
Thus, high values of $SR(I)$ indicate that $I$ is more frequent in $\mathcal{D}_{FPT}$.
Finally, we use a two-sided Fisher’s exact test to verify that the observed association between itemset presence and the FPT labels is statistically significant (\ie unlikely to have arisen by chance). The results confirm that the vast majority of retained itemsets exhibit a significant association with the FPT label. Specifically, of the 456 mined itemsets, 435 (95\%) are statistically significant, indicating that the observed enrichment of these patterns is not due to random variation.

This scoring step prioritizes itemsets that are (1) sufficiently prevalent to be robust and (2) disproportionately more prevalent in FPT-related traffic (\ie high support ratio), while deprioritizing patterns that occur frequently in both classes.
We focus only on itemsets with $\mathrm{SR}(I) > 1$, as these are more frequent in tracking traffic and relevant for rule generation. Out of 456 itemsets, 242 meet this criterion. They show a wide spread in support ratios, with a mean of 422.8, a median of 31.9, and a maximum of 7166.98. The high maximum and standard deviation (979.2) are due to top itemsets that occur frequently in $\mathcal{D}_{FPT}$ but are absent $\mathcal{D}_{Other}$. In contrast, the 214 itemsets with $\mathrm{SR}(I) < 1$ are more common in $\mathcal{D}_{Other}$ and are thus excluded. %
We select a threshold of $\mathrm{SR}(I)\geq 10$, meaning that an itemset is at least ten times more likely to occur in $\mathcal{D}_{FPT}$ than in $\mathcal{D}_{Other}$. This threshold was chosen to retain patterns that are strongly enriched in tracking traffic while avoiding overly restrictive filtering that would discard informative but less extreme itemsets. %
This threshold retains patterns strongly enriched in tracking traffic while avoiding overly restrictive filtering.

\subsection{Building FPT Blocking Rules}
To operationalize the statistically enriched itemsets from \cref{sec:sst_scoring} as deployable countermeasures, we automatically translate retained query-key combinations into Adblock Plus (ABP)–compatible URL filter rules (\ie rules that can be included into popular blocking lists like EasyList). This translation step is required because frequent itemsets are unordered sets of keys, whereas HTTP requests encode query parameters in an order that may vary across implementations.

\paragraph{Rule Generation Approach}
Frequent itemsets are inherently unordered, and the same set of query keys may appear in different orders in real traffic. To avoid emitting one rule per key permutation, we construct a single ABP-compatible regex that enforces set membership using positive lookaheads. Concretely, for an itemset with keys $k_1, ..., k_m$, we emit a rule that (1) anchors at a query boundary and (2) requires each key to occur somewhere in the query string (before the fragment delimiter). An example for the itemset \texttt{\{if, r, v\}} is:
{\footnotesize
\verb|/[\?&](?=[^#]*\bif=)(?=[^#]*\br=)(?=[^#]*\bv=)[^#]*/|
}
The prefix \verb|/[\?&]| anchors the match at the beginning of the query string (after \verb|?|) or at a
parameter boundary (after \verb|&|). Each lookahead, e.g., \verb|(?=[^#]*\bif=)|, enforces that the
corresponding key occurs somewhere in the query component before the fragment delimiter \verb|#|.
Because lookaheads do not consume characters, the required keys may appear in \emph{any} order and may be
separated by arbitrary intervening parameters. Finally, \verb|[^#]*| consumes the remainder of the query
string while ensuring the match does not cross into the fragment part of the URL.
For single-key itemsets, we emit the simpler rule \verb|/[?&]k=/| without lookaheads. This procedure provides a systematic bridge from validated co-occurrence patterns to compact blocking rules that can be applied to existing URL-filtering methods.

\paragraph{Rule Evaluation}
Following the presented approach, we derived 181 filter list rules to protect against FPT.
To evaluate these rules, we test them against our validation datasets $\mathcal{V}_{FPT}$ and $\mathcal{V}_{Other}$.
If our rules match any of the URLs in $\mathcal{V}_{FPT}$, we assume that it would have blocked an FPT-related request, while blocking a request in $\mathcal{V}_{Other}$ is most likely a false positive that could lead to page breakage. 
We find that the rules correctly classified 5,482 (63\%) of the URLs in $\mathcal{V}_{FPT}$ while EasyList and EasyPrivacy flagged none of these requests.
Thus, our approach provides reliable protection against FPT with only a few filter rules ($\approx 0.2\%$ of the used EasyList).
When applying the rules to $\mathcal{V}_{Other}$, we find that the rules would have blocked 260 (3\%) URLs.
To better understand the potential impact of falsely blocking these requests, we perform a visual page-breakage analysis.

\subsection{Visual Page Breakage Analysis}
To evaluate the usability of our blocking rules in a real-world scenario. We perform a \emph{visual} page breakage analysis~\cite{smith2022blockedbrokenautomaticallydetecting,boettger.2025,cookiegraph} on the top 2,300 URLs from the Tranco list\footnote{\url{https://tranco-list.eu/list/X4N2N/1000000}} %
using the created set of 181 blocking rules. Therefore, we configured two OpenWPM instances to visit each site's landing page, wait until the page loads (or 30 seconds), and take a screenshot of the full page. One crawling instance uses the AdBlock Plus extension~\cite{adblockplus2024}, which loads our generated rules rather than the standard blocking lists used by the tool. The other instance is a vanilla OpenWPM instance serving as a baseline for the page-breakage analysis. %
We started both instances on two identically configured virtual machines at the same time. 

To identify visual page breakage, two authors compare screenshots of a site to look for clues that a page has stopped working. For this coding the authors used four different codes: (1) \emph{Working}, indicating that the page seems to work as usual; (2) \emph{Broken}, indicating that the page stopped working; (3) \emph{Blocked}, indicating that the page is showing a warning message that an ad blocker is being used that should be deactivated to continue; (4) \emph{Missing}, indicating that at least one screenshot of a page is missing (\ie one of the instances could not access the page); (5) \emph{Denied}, indicating an error (\eg ``access denied`` or bot detection check) on the page.

Of the analyzed websites, 668 (29\%) could not be analyzed in both instances (``Missing''), 135 (5.8\% denied the access to the page (``Denied``). 32 (1.4\% page detected the usage of an ad blocker (``Blocked''), and 1,464 (63.7\%) seem to work as expected (``Working''). 
Consequently, 1 (0.04\%) of the page did not work properly after applying the blocking rules, suggesting that the approach is valid for building filter rules.
It was expected that only a few pages would identify the use of an ad blocker, as pages typically do so by querying a known tracking URL to see if the request is being blocked. Our rules will not block these requests.

\section{Related Work}
\label{sec:rw}
\label{sec:server-side-tracking}
El Farihi \etAl found SST nearly as effective as client-side tracking on Meta, but with a 40\% false-match rate~\cite{elfraihi:hal-04665102}. Alegría et al. analyze GTM tags across the million most popular sites and identify violations of GTM's permission system~\cite{Alegri2025}. Our work situates GTM within the broader FPT/SST ecosystem rather than studying it in isolation.
In a further approach, Munir \etAl present CookieGraph~\cite{cookiegraph}. 
In their work, they perform large-scale measurements to collect first- and third-party cookies, labeling them with a combination of filter lists and an ML-based approach to identify tracking requests. CookieGraph is trained with labeled data (supervised learning) to detect and block first-party tracking cookies. We aim to provide a broad overview of FPT, including its ecosystem. While CookieGraph mainly focuses on detecting first-party cookies. %
Nikkhah Bahrami \etAl propose a browser-based mechanism to block unauthorized cross-domain cookie usages~\cite{NikkhahBahrami2025}.
We dig deeper into FPT by not limiting ourselves to a specific platform or technique, but by providing a generalizable approach to identify them on the Web and by analyzing their ecosystem. Further, we provide a first approach to building privacy-enhancing technologies that protect against FPT.

Fouad \etAl focus on the non-compliant practices of SST in the context of the GDPR and ePrivacy Directive~\cite{Fouad2024} and use a cloaking-based approach to identify potential SST activities. %
In our approach, we avoid cloaking techniques and instead derive a rule-based detection approach.
The Web Almanac 2024 also reports a shift from third-party to first-party cookie usage. Major providers (\eg Meta) largely disappeared from third-party cookie settings in 2024~\cite{WebAlmanac2024}, while their first-party cookie settings increased~\cite{WebAlmanac.2024.Privacy}. %
Our results extend these findings by providing a comprehensive view of the FPT ecosystem and related trends.

\section{Conclusion}
Our large-scale measurement study, incorporating two vantage points and four measurement profiles, demonstrates that first-party tracking is a growing phenomenon increasingly shifting into the first-party context. This shift poses several downsides for users, including reduced transparency, diminished control, and limited ability to block such tracking, as current filter lists are not optimized for first-party resources. The results show that, in terms of source code, tracking scripts sharing the same cookies operate in both contexts, indicating that the FPT ecosystem mirrors its third-party counterpart and is dominated by the same well-known players (\eg Google and Meta). Based on the shared use of cookies and tracking scripts, we construct a network graph showing that the FPT ecosystem is highly interconnected, with dense co-deployment of cookies across clusters. To address the protection gap, we automatically derive filter rules tailored to FPT, demonstrating that they outperform current filter lists in the first-party context, underscoring the need for more effective blocking mechanisms. In conclusion, FPT poses an emerging privacy threat on the Web for which existing defenses remain inadequate.

\medskip
\noindent\textbf{Limitations and Future Work.} Our analysis is limited to client-side indicators of FPT. We cannot assess site backends, so we cannot draw conclusions about server-side data flows. Future work could explore server-side mechanisms by deploying trackers in controlled environments to understand their backend logic. We further restrict our scope to cookie-based tracking within JavaScript, excluding techniques such as tracking pixels, which we consider part of future work.

\section*{Acknowledgments}
The authors gratefully acknowledge funding from the \emph{German Federal Ministry of Education and Research} (grants 16KIS1629 ``UbiTrans'') and the \emph{German Federal Ministry for Housing, Urban Development and Building}. We also thank Google for the generous funding that supported this research.

\bibliographystyle{ACM-Reference-Format}
\bibliography{bibliography}

@string{isoc-ndss       = "Symposium on Network and Distributed System Security"}

@string{acsac           = "Anual Computer Security Applications Conference"}

@string{www             = "International Conference on World Wide Web"}

@string{acm-imc         = "ACM SIGCOMM Internet Measurement Conference"}

@string{pets            = "Proceedings on Privacy Enhancing Technologies"}

@string{pro             = "Proceedings of the "}

@string{chi             = "ACM SIGCHI Conference on Human Factors in Computing Systems"}

@string{pam             = "Conference on Passive and Active Measurement"}

@string{tma             = "Network Traffic Measurement and Analysis Conference"}

@inproceedings{nguyen_consent_2022,
    author = {Nguyen, Trung Tin and Backes, Michael and Stock, Ben},
    title = {{Freely Given Consent? Studying Consent Notice of Third-Party Tracking and Its Violations of GDPR in Android Apps}},
    booktitle = {Proceedings of the 2022 ACM SIGSAC Conference on Computer and Communications Security},
    series = {CCS '22},
    year = {2022},
    pages = {2369-2383},
    numpages = {15},
    howpublished = {\url{https://doi.org/10.1145/3548606.3560564}}  ,
    doi = {10.1145/3548606.3560564},
    publisher = {ACM},
    address = {New York, NY, USA},
}

@inproceedings{gonzalez_cookies_2017,
    author={Gonzalez, Roberto and Jiang, Lili and Ahmed, Mohamed and Marciel, Miriam and Cuevas, Ruben and Metwalley, Hassan and Niccolini, Saverio},
    title = {{The Cookie Recipe: Untangling the Use of Cookies in the Wild}}, 
    booktitle = {Proceedings of the 1st Network Traffic Measurement and Analysis Conference}, 
    series = {TMA 2017},
    year = {2017},
    articleno = {1},
    numpages = {9},
    howpublished = {\url{https://doi.org/10.23919/TMA.2017.8002896}} ,
    doi = {10.23919/TMA.2017.8002896},
    publisher = {IEEE},
    address = {Washington, DC, USA},
}

@inproceedings{Acar.tracking.2014, 
    author = {Acar, Gunes and Eubank, Christian and Englehardt, Steven and Juarez, Marc and Narayanan, Arvind and Diaz, Claudia},
    title = {The Web Never Forgets: Persistent Tracking Mechanisms in the Wild},
    booktitle = {Proceedings of the 2014 ACM SIGSAC Conference on Computer and Communications Security},
    series = {CCS '14},
    year = {2014},
    location = {},    
    pages = {674–689},
    numpages = {16},
    howpublished = {\url{https://doi.org/10.1145/2660267.2660347}} ,
    doi = {10.1145/2660267.2660347},
    publisher = {Association for Computing Machinery},
    address = {New York, NY, USA},
    
}

@inproceedings{urban_adnetworks_2020,
    author = {Urban, Tobias and Tatang, Dennis and Degeling, Martin and Holz, Thorsten and Pohlmann, Norbert},
    title = {{Measuring the Impact of the GDPR on Data Sharing in Ad Networks}},
    booktitle = {Proceedings of the 15th ACM Asia Conference on Computer and Communications Security},    
    series = {ASIA CCS '20},
    year = {2020},
    pages = {222–235},
    numpages = {14},
    howpublished = {\url{https://doi.org/10.1145/3320269.3372194}}  ,
    doi = {10.1145/3320269.3372194},
    publisher = {ACM},
    address = {New York, NY, USA},
}

@misc{Cookiepedia.2024,
    author = {{OneTrust LLC}},
    title = {{Cookiepedia -- All You Need to Know About Cookies}},
    howpublished = {\url{https://cookiepedia.co.uk/}},
    year = {2026},
}

@ARTICLE{Liu.Opt.out.2024,
  title     = "Opted out, yet tracked: Are regulations enough to protect your privacy?",
  author    = "Liu, Zengrui and Iqbal, Umar and Saxena, Nitesh",
  journal   = "Proc. Priv. Enhancing Technol.",
  publisher = "Privacy Enhancing Technologies Symposium Advisory Board",
  volume    =  2024,
  number    =  1,
  pages     = "280--299",
  month     =  jan,
  year      =  2024
}

@ARTICLE{Rasaii.cmp.2025,
  title     = "Intractable cookie crumbs: Unveiling the nexus of stateful
               banner interaction and tracking cookies",
  author    = "Rasaii, Ali and Dao, Ha and Feldmann, Anja and Javid,
               Mohammadmahdi and Gasser, Oliver and Gosain, Devashish",
  journal   = "Proc. Priv. Enhancing Technol.",
  publisher = "Privacy Enhancing Technologies Symposium Advisory Board",
  volume    =  2025,
  number    =  4,
  pages     = "429--445",
  month     =  oct,
  year      =  2025
}

@inproceedings{Nouwens.cmp.2025,
    author = {Nouwens, Midas and Kristensen, Janus Bager and Maalt, Kristjan and Bagge, Rolf},
    title = {A Cross-Country Analysis of GDPR Cookie Banners and Flexible Methods For Scraping Them},
    year = {2025},
    isbn = {9798400713941},
    publisher = {Association for Computing Machinery},
    address = {New York, NY, USA},
    doi = {10.1145/3706598.3713648},
    booktitle = {Proceedings of the 2025 CHI Conference on Human Factors in Computing Systems},
    articleno = {858},
    numpages = {28},
    location = {
    },
    series = {CHI '25}
}

@InProceedings{Santos.CMP.2021,
    author="Santos, Cristiana
    and Nouwens, Midas
    and Toth, Michael
    and Bielova, Nataliia
    and Roca, Vincent",
    title="Consent Management Platforms Under the GDPR: Processors and/or Controllers?",
    booktitle="Privacy Technologies and Policy",
    year="2021",
    publisher="Springer International Publishing",
    address="Cham",
    pages="47--69",
    isbn="978-3-030-76663-4"
}

@misc{Cookiedatabase.2026,
    author = {{SIDN fonds}},
    title = {{Cookiedatabase -- Understanding tracking with Cookiedatabase.org}},
    howpublished = {\url{https://www.cookiedatabase.org/}},
    year = {2026},
}

@misc{abck.cookie.2026,
    author = {{Outis Nemo Ltd. }},
    title = {Akamai \_abck Cookie},
    howpublished = {\url{https://kameleo.io/glossary/akamai-abck-cookie/}},
    year = {2026},
}

@misc{abck.cookie.comp.2026,
    author = {{Captain Compliance}},
    title = {\_abck Cookie: What Does it Mean?},
    howpublished = {\url{https://captaincompliance.com/education/_abck/}},
    year = {2026},
}

@inproceedings{utz_consent_2019,
    author = {Christine Utz and Martin Degeling and Sascha Fahl and Florian Schaub and Thorsten Holz},
    title = {{(Un)informed Consent: Studying GDPR Consent Notices in the Field}},
    booktitle = {Proceedings of the 2019 ACM SIGSAC Conference on Computer and Communications Security},
    series = {CCS '19},
    year = {2019},
    pages = {973–990},
    numpages = {18},
    howpublished = {\url{https://doi.org/doi/10.1145/3319535.3354212}},
    doi = {10.1145/3319535.3354212},
    publisher = {ACM},
    address = {New York, NY, USA},
}

@misc{EasyList.2023,
    author = {{EasyList authors}, The},
    title = {{EasyList}},
    howpublished = {\url{https://easylist.to/easylist/easylist.txt}},
    year = {2023},
    month = mar,
    day = 23,
    note = {Used list from March 23, 2023; version 202303230338}
}

@misc{EasyListPrivacy.2024,
    author = {{EasyList authors}, The},
    title = {{EasyPrivacy}},
    howpublished = {\url{https://easylist.to/easylist/easyprivacy.txt}},
    year = {2024},
    month = jul,
    day = 22,
    note = {Used list from July 22, 2024; version 202407221302}
}

@article{fouad2020,
    author = {Imane Fouad and Nataliia Bielova and Arnaud Legout and Natasa Sarafijanovic-Djukic},
    title = {{Missed by Filter Lists: Detecting Unknown Third-Party Trackers with Invisible Pixels}},
    journal = {Proceedings on Privacy Enhancing Technologies},
    issue_date = {Apri 2020},
    volume = {2020},
    number = {2},
    month = apr,
    year = {2020},
    pages = {499-518},
    numpages = {20},
    howpublished = {\url{https://doi.org/10.2478/popets-2020-0038}},
    doi = {10.2478/popets-2020-0038},
    publisher = {Sciendo},
    address = {Warsaw, Poland},
}

@article{Demir.banner.24,
    author = {Demir, Nurullah 
        and Urban, Tobias
        and Wressnegger, Christian
        and Pohlmann, Norbert},
    title = {{A Large-Scale Study of Cookie Banner Interaction Tools and Their Impact on Users' Privacy}},
    journal = {Proceedings on Privacy Enhancing Technologies},
    issue_date = {2022},
    volume = {2024},
    number = {1},
    year = {2024},
    pages = {5--20},
    numpages = {16},
    howpublished = {\url{https://doi.org/10.56553/popets-2024-0002}},
    doi = {10.56553/popets-2024-0002},
    publisher = {Self-published},
    address = {},
}

@inproceedings{hagberg_networkx_2008,
    author = {Aric A. Hagberg and Daniel A. Schult and Pieter J. Swart},
    title = {{Exploring Network Structure, Dynamics, and Function using NetworkX}},
    booktitle = {Proceedings of the 7th Python in Science Conference},
    series = {SciPy 2008},
    year = {2008},
    pages = {11 - 15},
    numpages = {5},
    howpublished = {\url{http://conference.scipy.org.s3-website-us-east-1.amazonaws.com/proceedings/SciPy2008/paper_2/}},
    doi = {},
    editor = {Ga\"el Varoquaux and Travis Vaught and Jarrod Millman},
    publisher = {},
    address = {},
}

@inproceedings{elfraihi:hal-04665102,
  TITLE = {{Client-side and Server-side Tracking on Meta: Effectiveness and Accuracy}},
  AUTHOR = {El fraihi, Asmaa and Amieur, Nardjes and Rudametkin, Walter and Goga, Oana},
  howpublished = {\url{https://hal.science/hal-04665102}},
  BOOKTITLE = {{PETS 2024 - 24th Privacy Enhancing Technologies Symposium}},
  ADDRESS = {Bristol, United Kingdom},
  VOLUME = {2024},
  PAGES = {431-445},
  YEAR = {2024},
  MONTH = Jul,
  DOI = {10.56553/popets-2024-0086},
  HAL_ID = {hal-04665102},
  HAL_VERSION = {v1},
}

@ARTICLE{9403411,
  author={Dao, Ha and Mazel, Johan and Fukuda, Kensuke},
  journal={IEEE Transactions on Network and Service Management}, 
  title={CNAME Cloaking-Based Tracking on the Web: Characterization, Detection, and Protection}, 
  year={2021},
  volume={18},
  number={3},
  pages={3873-3888},
  doi={10.1109/TNSM.2021.3072874}}

@misc{google_sst_fundamentals,
  author       = {Google Developers},
  title        = {What is SST? - SST Fundamentals},
  howpublished = {\url{https://developers.google.com/tag-platform/learn/sst-fundamentals/2-what-is-sst}},
  year         = 2024
}

@misc{google_server_side_tagging,
  author       = {Ben Fisher},
  title        = {Improve performance and security with Server-Side Tagging},
  howpublished = {\url{https://blog.google/products/marketingplatform/360/improve-performance-and-security-server-side-tagging/}},
  year         = 2020,
  month        = aug,
  publisher    = {Google Marketing Platform}
}

@article{demir.2022,
  author       = {Nurullah Demir and
                  Daniel Theis and
                  Tobias Urban and
                  Norbert Pohlmann},
  title        = {Towards Understanding First-Party Cookie Tracking in the Field},
  journal      = {CoRR},
  volume       = {abs/2202.01498},
  year         = {2022},
  howpublished          = {\url{https://arxiv.org/abs/2202.01498}},
  eprinttype    = {arXiv},
  eprint       = {2202.01498},
  bibhowpublished       = {https://dblp.org/rec/journals/corr/abs-2202-01498.bib},
  bibsource    = {dblp computer science bibliography, https://dblp.org}
}

@misc{Demir_MultiCrawl,
    author = {{Demir, Nurullah}},
    title = {{MultiCrawl}},
    howpublished = {\url{https://github.com/nrllh/multicrawl}},
    year = {2023},
    note = {Used version: MultiCrawl (v0.1.2)}
}

@article{ratcliff1988,
  author  = {Ratcliff, J.W. and Metzener, D.E.},
  title   = {Pattern Matching: The Gestalt Approach},
  journal = {Dr Dobbs Journal},
  volume  = {13},
  number  = {7},
  year    = {1988},
}

@inproceedings{Khan.2018,
 author = {Khan, Mohammad Taha and DeBlasio, Joe and Voelker, Geoffrey M. and Snoeren, Alex C. and Kanich, Chris and Vallina-Rodriguez, Narseo},
 title = {An Empirical Analysis of the Commercial VPN Ecosystem},
 booktitle = pro # "2018 " # acm-imc,
 series = {IMC'18},
 year = {2018},
 numpages={15},
 publisher = {ACMPress},
 address = {New York, NY, USA},
}

@inproceedings{10.1145/2736277.2741679,
author = {Englehardt, Steven and Reisman, Dillon and Eubank, Christian and Zimmerman, Peter and Mayer, Jonathan and Narayanan, Arvind and Felten, Edward W.},
title = {Cookies That Give You Away: The Surveillance Implications of Web Tracking},
year = {2015},
isbn = {9781450334693},
publisher = {International World Wide Web Conferences Steering Committee},
address = {Republic and Canton of Geneva, CHE},
howpublished = {\url{https://doi.org/10.1145/2736277.2741679}}              ,
doi = {10.1145/2736277.2741679},
booktitle = {Proceedings of the 24th International Conference on World Wide Web},
pages = {289–299},
numpages = {11},
location = {Florence, Italy},
series = {WWW '15}
}

@article{Koop2020InDepthEO,
  title={In-Depth Evaluation of Redirect Tracking and Link Usage},
  author={Martin Koop and Erik Tews and Stefan Katzenbeisser},
  journal={Proceedings on Privacy Enhancing Technologies},
  year={2020},
  volume={2020},
  pages={394 - 413},
  howpublished={\url{https://api.semanticscholar.org/CorpusID:220531844}}
}

@misc{WebAlmanac2024,
  author       = {{Web Almanac}},
  title        = {The 2024 Web Almanac: A Comprehensive Report on the State of the Web},
  year         = {2024},
  howpublished = {\url{https://almanac.httparchive.org/en/2024/}}
}

@inbook{WebAlmanac.2024.Privacy,
author = "Vekaria, Yash and Standaert, Benjamin and Ostapenko, Max and Amjad, Abdul Haddi and Dimova, Yana and Munir, Shaoor and Böttger, Chris and Iqbal, Umar and Fernandez-de-Retana, Alberto and Pollard, Barry",
title = "Privacy",
booktitle = "The 2024 Web Almanac",
chapter = 12,
publisher = "HTTP Archive",
year = "2024",
language = "English",
doi = "10.5281/zenodo.14261510",
howpublished = "https://almanac.httparchive.org/en/2024/privacy"
}

@inproceedings{cookiegraph,
author = {Munir, Shaoor and Siby, Sandra and Iqbal, Umar and Englehardt, Steven and Shafiq, Zubair and Troncoso, Carmela},
title = {CookieGraph: Understanding and Detecting First-Party Tracking Cookies},
year = {2023},
isbn = {9798400700507},
publisher = {Association for Computing Machinery},
address = {New York, NY, USA},
howpublished = {https://doi.org/10.1145/3576915.3616586}       ,
doi = {10.1145/3576915.3616586},
booktitle = {Proceedings of the 2023 ACM SIGSAC Conference on Computer and Communications Security},
pages = {3490–3504},
numpages = {15},
location = {Copenhagen, Denmark},
series = {CCS '23}
}

@article{Hamming1950,
  author    = {Hamming, Richard W.},
  title     = {Error Detecting and Error Correcting Codes},
  journal   = {Bell System Technical Journal},
  volume    = {29},
  number    = {2},
  pages     = {147--160},
  year      = {1950},
  doi       = {https://doi.org/10.1002/j.1538-7305.1950.tb00463.x}       
}

@article{boettger.2025,
  title = {Understanding Regional Filter Lists: Efficacy and Impact},
  volume = {2025},
  ISSN = {2299-0984},
  howpublished = {\url{http://dx.doi.org/10.56553/popets-2025-0063}}   ,
  DOI = {10.56553/popets-2025-0063},
  number = {2},
  series = {PoPETS}, 
  journal = {Proceedings on Privacy Enhancing Technologies},
  publisher = {Privacy Enhancing Technologies Symposium Advisory Board},
  author = {B\"{o}ttger,  Christian and Demir,  Nurullah and H\"{o}rnemann,  Jan and Acharya,  Bhupendra and Pohlmann,  Norbert and Holz,  Thorsten and Grosse-Kampmann,  Matteo and Urban,  Tobias},
  year = {2025},
  month = apr,
  pages = {309–325}
}

@InProceedings{Demir.updates.2021,
    title = {{Our (in)Secure Web: Understanding Update Behavior of Websites and Its Impact on Security}},
    booktitle = pam,
    series = {PAM},
    author = {Demir, Nurullah 
        and Urban, Tobias
        and Wittek, Kevin
        and Pohlmann, Norbert},
    publisher = {Springer-Verlag},
    address = {Berlin, Heidelberg},
    pages = {76–92},
    year = {2021},
    isbn = {978-3-030-72581-5 },
    doi = {10.1007/978-3-030-72582-2_5 },
}

@inproceedings{Demir_rep_2022,
    author = {Demir, Nurullah and Gro\ss{}e-Kampmann, Matteo and Urban, Tobias and Wressnegger, Christian and Holz, Thorsten and Pohlmann, Norbert},
    title = {Reproducibility and Replicability of Web Measurement Studies},
    year = {2022},
    isbn = {9781450390965},
    publisher = {Association for Computing Machinery},
    address = {New York, NY, USA},
    url = {https://doi.org/10.1145/3485447.3512214},
    doi = {10.1145/3485447.3512214},
    booktitle = {Proceedings of the ACM Web Conference 2022},
    pages = {533–544},
    numpages = {12},
    location = {Virtual Event, Lyon, France},
    series = {WWW}
}

@InProceedings{Calzavara.Headers.2021,
    author = {Calzavara, Stefano
        and Urban, Tobias 
        and Tatang, Dennis 
        and Steffens, Marius 
        and Stock, Ben},
    title = {{Reining in the Web's Inconsistencies with Site Policy}},
    booktitle = isoc-ndss,
    series = {NDSS},
    year = {2021},
    doi = {10.14722/ndss.2021.23091},
}

@misc{PublicSuffixList.2025,
    author = {{Mozilla Foundation}},
    title = {{Public Suffix List}},
    howpublished = {\url{https://publicsuffix.org/list/public_suffix_list.dat}},
    year = {2025}
}

@InProceedings{Urban.WWW.2020,
    author = {Urban, Tobias 
        and Degeling, Martin 
        and Holz, Thorsten 
        and Pohlmann, Norbert},
    title = {{Beyond the Front Page: Measuring Third Party Dynamics in the Field}},
    year = {2020},
    booktitle = www,
publisher = {Association for Computing Machinery},
address = {New York, NY, USA},
pages = {1275–1286},
    series = {TheWebConf},
    doi = {10.1145/3366423.3380203 },
}

@inproceedings{LePochat2019,
   author = "{Le Pochat}, Victor and {Van Goethem}, Tom and Tajalizadehkhoob, Samaneh and Korczy\'{n}ski, Maciej and Joosen, Wouter",
    title = "Tranco: A Research-Oriented Top Sites Ranking Hardened Against Manipulation",
booktitle = {Proceedings of the 26th Annual Network and Distributed System Security Symposium},
   series = {NDSS 2019},
     year = 2019,
    month = feb,
      doi = {10.14722/ndss.2019.23386 },
}

@inproceedings{urban.2019,
  author       = {Tobias Urban and
                  Martin Degeling and
                  Thorsten Holz and
                  Norbert Pohlmann},
  editor       = {David M. Balenson},
  title        = {"Your hashed {IP} address: Ubuntu.": perspectives on transparency tools for online advertising},
  booktitle    = {Proceedings of the 35th Annual Computer Security Applications Conference},
  series       = {ACSAC},
  pages        = {702--717},
  publisher    = {{ACM}},
  year         = {2019},
  howpublished          = {https://doi.org/10.1145/3359789.3359798},
  doi          = {10.1145/3359789.3359798},
}

@inproceedings{demir_2023,
    author = {Demir, Nurullah and H\"{o}rnemann, Jan and Gro\ss{}e-Kampmann, Matteo and Urban, Tobias and Pohlmann, Norbert and Holz, Thorsten and Wressnegger, Christian},
    title = {On the Similarity of Web Measurements Under Different Experimental Setups},
    year = {2023},
    isbn = {9798400703829 },
    publisher = {Association for Computing Machinery},
    address = {New York, NY, USA},
    howpublished = {https://doi.org/10.1145/3618257.3624795                     }        , 
    booktitle = {Proceedings of the 2023 ACM on Internet Measurement Conference},
    pages = {356–369},
    numpages = {14},
    location = {Montreal QC, Canada},
    series = {IMC '23}
}

@article{Prince.2024,
    author = {Prince, Christine and Omrani, Nessrine and Schiavone, Francesco},
    title = {Online privacy literacy and users' information privacy empowerment: the case of GDPR in Europe},
    journal = {Information Technology \& People},
    volume = {37},
    number = {8},
    pages = {1-24},
    year = {2024},
    month = {01},
    issn = {0959-3845},
    doi = {10.1108/ITP-05-2023-0467},
    howpublished = {https://doi.org/10.1108/ITP-05-2023-0467},
    eprint = {https://www.emerald.com/itp/article-pdf/37/8/1/9465373/itp-05-2023-0467.pdf},
}

@inproceedings{Sood.11,
author = {Sood, Sadhan and Loguinov, Dmitri},
title = {Probabilistic near-duplicate detection using simhash},
year = {2011},
isbn = {9781450307178},
publisher = {Association for Computing Machinery},
address = {New York, NY, USA},
howpublished = {https://doi.org/10.1145/2063576.2063737},
doi = {10.1145/2063576.2063737},
booktitle = {Proceedings of the 20th ACM International Conference on Information and Knowledge Management},
pages = {1117–1126},
numpages = {10},
location = {Glasgow, Scotland, UK},
series = {CIKM '11}
}

@article{Kretschmer.2021,
    author = {Kretschmer, Michael and Pennekamp, Jan and Wehrle, Klaus},
    title = {Cookie Banners and Privacy Policies: Measuring the Impact of the GDPR on the Web},
    year = {2021},
    issue_date = {November 2021},
    publisher = {Association for Computing Machinery},
    address = {New York, NY, USA},
    volume = {15},
    number = {4},
    issn = {1559-1131},
    howpublished = {https://doi.org/10.1145/3466722},
    doi = {10.1145/3466722},
    journal = {ACM Trans. Web},
    month = jul,
    articleno = {20},
    numpages = {42}
}

@misc{iab.2024,
  author = {{PricewaterhouseCoopers} and {Interactive Advertising Bureau}},
  title = {Internet Advertising Revenue Report},
  howpublished = {\url{https://www.iab.com/wp-content/uploads/2025/04/IAB\_PwC-Internet-Ad-Revenue-Report-Full-Year-2024.pdf}},
  year={2024},
}

@misc{consentomatic2021,
  author       = {Jensen, Anders and Kristensen, Mads Mølgaard and Schultz, Thomas Tønnesen and Hildebrandt, Thomas and Andersen, Michael},
  title        = {{Consent-O-Matic}},
  year         = {2021},
  publisher    = {CAVI, Aarhus University},
  howpublished          = {\url{https://github.com/cavi-au/Consent-O-Matic}},
  note         = {Browser extension for automatic GDPR consent management. Accessed: 2025-04-16}
}

@article{JARVINEN2015117,
title = {The use of Web analytics for digital marketing performance measurement},
journal = {Industrial Marketing Management},
volume = {50},
pages = {117-127},
year = {2015},
issn = {0019-8501},
doi = {https://doi.org/10.1016/j.indmarman.2015.04.009},
url = {https://www.sciencedirect.com/science/article/pii/S001985011500139X},
author = {Joel Järvinen and Heikki Karjaluoto},
}

@misc{nordvpn,
  author       = {{NordVPN}},
  title        = {NordVPN Official Website},
  year         = {2025},
  howpublished          = {https://nordvpn.com}
}

@misc{Google2020ServerSideTagging,
  author    = {{Google}},
  title     = {Improve performance and security with server-side tagging},
  year      = {2020},
  month     = aug,
  howpublished       = {\url{https://blog.google/products/marketingplatform/360/improve-performance-and-security-server-side-tagging/}}
}

@misc{Google2025GTMDocs,
  author    = {{Google Developers}},
  title     = {Introduction to server-side tagging in Google Tag Manager},
  year      = {2025},
  howpublished       = {https://developers.google.com/tag-platform/tag-manager/server-side/intro}
}

@misc{Meta2025SignalsGateway,
  author    = {{JENTIS}},
  title     = {Meta Signals Gateway: New strategic approach to server-side tracking},
  year      = {2025},
  month     = feb,
  howpublished       = {\url{https://www.jentis.com/en/article/blog-meta-signals-gateway/}}
}

@techreport{Jentis2024Whitepaper,
  author    = {{JENTIS}},
  title     = {How to find and implement the ideal server-side tracking solution},
  institution = {JENTIS GmbH},
  year      = {2024},
  month     = sep,
  howpublished       = {\url{https://www.jentis.com/wp-content/uploads/2024/09/JENTIS_White_Paper_How_to_find_and_implement_the_ideal_SST_solution_EN-compressed.pdf}}
}

@misc{tiktok_gtm_pixel_2025,
  author       = {{TikTok Inc.}},
  title        = {TikTok Pixel Template for Google Tag Manager},
  year         = {2025},
  howpublished = {\url{https://github.com/tiktok/gtm-template-pixel}},
  note         = {Official GitHub repository providing the TikTok Pixel Tag template for Google Tag Manager. Accessed: 2025-10-21.},
  organization = {TikTok Pte. Ltd.}
}

@misc{1e0ng_simhash,
  author       = {{1e0ng on GitHub}},
  title        = {SimHash: a Python implementation of the Simhash algorithm},
  year         = {2013},
  howpublished = {\url{https://github.com/1e0ng/simhash}}
}

@inproceedings{charikar2002simhash,
  author    = {Moses S. Charikar},
  title     = {Similarity Estimation Techniques from Rounding Algorithms},
  booktitle = {Proceedings of the 34th Annual ACM Symposium on Theory of Computing (STOC '02)},
  year      = {2002},
  pages     = {380--388},
  publisher = {Association for Computing Machinery},
  address   = {New York, NY, USA},
  doi       = {10.1145/509907.509965}
}

@article{lex2014upset,
  author    = {Alexander Lex and Nils Gehlenborg and Hendrik Strobelt and Romain Vuillemot and Hanspeter Pfister},
  title     = {UpSet: Visualization of Intersecting Sets},
  journal   = {IEEE Transactions on Visualization and Computer Graphics},
  year      = {2014},
  volume    = {20},
  number    = {12},
  pages     = {1983--1992},
  doi       = {10.1109/TVCG.2014.2346248}
}

@inproceedings{Binns2018,
author = {Binns, Reuben and Lyngs, Ulrik and Van Kleek, Max and Zhao, Jun and Libert, Timothy and Shadbolt, Nigel},
title = {Third Party Tracking in the Mobile Ecosystem},
year = {2018},
isbn = {9781450355636},
publisher = {Association for Computing Machinery},
address = {New York, NY, USA},
howpublished = {https://doi.org/10.1145/3201064.3201089} , 
booktitle = {Proceedings of the 10th ACM Conference on Web Science},
pages = {23–31},
numpages = {9},
location = {Amsterdam, Netherlands},
series = {WebSci '18}
}

@misc{falahrastegar2014anatomythirdpartywebtracking,
      title={Anatomy of the Third-Party Web Tracking Ecosystem}, 
      author={Marjan Falahrastegar and Hamed Haddadi and Steve Uhlig and Richard Mortier},
      year={2014},
      eprint={1409.1066},
      archivePrefix={arXiv},
      primaryClass={cs.SI},
      howpublished={\url{https://arxiv.org/abs/1409.1066}}, 
}

@phdthesis{Falahrastegar2017,
  author       = {Marjan Falahrastegar},
  title        = {The complex third-party tracking ecosystem: a multi­-dimensional perspective},
  school       = {Queen Mary, University of London},
  year         = {2017},
  howpublished          = {https://qmro.qmul.ac.uk/xmlui/handle/123456789/25818},
  note         = {PhD thesis}
}

@inproceedings{Razaghpanah2018,
  author       = {Abbas Razaghpanah and Rishab Nithyanand and Narseo Vallina-Rodriguez and Srikanth Sundaresan and Mark Allman and Christian Kreibich and Phillipa Gill},
  title        = {Apps, Trackers, Privacy, and Regulators: A Global Study of the Mobile Tracking Ecosystem},
  booktitle    = {Proceedings of the 2018 Network and Distributed System Security Symposium (NDSS)},
  year         = {2018},
  howpublished          = {https://dspace.networks.imdea.org/handle/20.500.12761/507},
  note         = {Paper}
}

@inproceedings{englehardt2016online,
  title={Online tracking: A 1-million-site measurement and analysis},
  author={Englehardt, Steven and Narayanan, Arvind},
  booktitle={Proceedings of the ACM SIGSAC Conference on Computer and Communications Security (CCS)},
  pages={1388--1401},
  year={2016},
  organization={ACM},
  doi={10.1145/2976749.2978313}
}

@misc{cookie_svweb.2026,
  author       = {{Cookiedatabase.org}},
  title        = {s\_v\_web\_id -- Summary},
  year         = {2026},
  howpublished = {\url{https://cookiedatabase.org/cookie/tiktok/s_v_web_id/}},
}

@misc{whotracksme,
    title={WhoTracks.Me: Shedding light on the opaque world of online tracking},
    author={Arjaldo Karaj and Sam Macbeth and Rémi Berson and Josep M. Pujol},
    year={2018},
    eprint={1804.08959},
    archivePrefix={arXiv},
    primaryClass={cs.CY}
}

@INPROCEEDINGS{SharifRox2011,
  author={Uddin, Md. Sharif and Roy, Chanchal K. and Schneider, Kevin A. and Hindle, Abram},
  booktitle={2011 18th Working Conference on Reverse Engineering}, 
  title={On the Effectiveness of Simhash for Detecting Near-Miss Clones in Large Scale Software Systems}, 
  year={2011},
  volume={},
  number={},
  pages={13-22},
  doi={10.1109/WCRE.2011.12}}

@inproceedings{SinghJain2007,
author = {Manku, Gurmeet Singh and Jain, Arvind and Das Sarma, Anish},
booktitle = {WWW 07: Proceedings of the 16th international conference on World Wide Web
},
title = {Detecting near-duplicates for web crawling},
year = {2007},
isbn = {9781595936547},
publisher = {Association for Computing Machinery},
address = {New York, NY, USA},
howpublished = {https://doi.org/10.1145/1242572.1242592}, 
pages = {141–150},
numpages = {10},
location = {Banff, Alberta, Canada},
series = {WWW '07}
}

@inproceedings{NikkhahBahrami2025,
author = {Nikkhah Bahrami, Pouneh and Fass, Aurore and Shafiq, Zubair},
title = {$CookieGuard:$ Characterizing and Isolating the First-Party Cookie Jar},
year = {2025},
isbn = {9798400718601},
publisher = {Association for Computing Machinery},
address = {New York, NY, USA},
howpublished = {https://doi.org/10.1145/3730567.3764490},
booktitle = {Proceedings of the 2025 ACM Internet Measurement Conference},
pages = {645–661},
numpages = {17},
location = {USA},
series = {IMC '25}
}

@inproceedings{Xuehui2021,
author = {Hu, Xuehui and Sastry, Nishanth and Mondal, Mainack},
title = {CCCC: Corralling Cookies into Categories with CookieMonster},
year = {2021},
isbn = {9781450383301},
publisher = {Association for Computing Machinery},
address = {New York, NY, USA},
howpublished = {\url{https://doi.org/10.1145/3447535.3462509}},
booktitle = {Proceedings of the 13th ACM Web Science Conference 2021},
pages = {234–242},
numpages = {9},
location = {Virtual Event, United Kingdom},
series = {WebSci '21}
}

@misc{cookie.isDatabase,
	author = {CookieHub},
	title = {{C}ookie {D}atabase --- cookie.is},
	howpublished = {\url{https://www.cookie.is/}},
	year = {2026}
}

@misc{rfc6265,
    publisher = {RFC Editor},
    doi =       {10.17487/RFC6265},
    howpublished =  {\url{https://datatracker.ietf.org/doc/html/rfc6265}},
    author =    {Adam Barth},
    title =     {RFC 6265: HTTP State Management Mechanism},
    pagetotal = 37,
    year =      2011,
    month =     apr,
}

@article{kollnig2021before,
  title={Before and after GDPR: Tracking in mobile apps},
  author={Kollnig, Konrad and Binns, Reuben and Van Kleek, Max and Lyngs,
               Ulrik and Zhao, Jun and Tinsman, Claudine and Shadbolt, Nigel},
  journal={Internet Policy Review},
  volume={10},
  number={4},
  year={2021},
  doi={10.14763/2021.4.1611},
  howpublished={\url{https://doi.org/10.14763/2021.4.1611}}
}

@misc{ghostery_trackerdb,
  author       = {{Ghostery, Inc.}},
  title        = {Ghostery TrackerDB},
  year         = 2023,
  howpublished          = {\url{https://github.com/ghostery/trackerdb}}
}

@misc{smith2022blockedbrokenautomaticallydetecting,
      title={Blocked or Broken? Automatically Detecting When Privacy Interventions Break Websites}, 
      author={Michael Smith and Peter Snyder and Moritz Haller and Benjamin Livshits and Deian Stefan and Hamed Haddadi},
      year={2022},
      eprint={2203.03528},
      archivePrefix={arXiv},
      primaryClass={cs.CR},
      howpublished={\url{https://arxiv.org/abs/2203.03528}}, 
}

@misc{adblockplus2024,
  title        = {Adblock Plus Extension, Version 4.32.2.1},
  author       = {{eyeo GmbH}},
  year         = {2024},
  version      = {4.32.2.1},
  howpublished          = {\url{https://adblockplus.org/}},
  note         = {Browser extension for ad blocking. Available for Chrome, Firefox, Edge and others.},
}

@misc{bonfils2025empiricalinquirysurveillancecapitalism,
      title={An Empirical Inquiry into Surveillance Capitalism: Web Tracking}, 
      author={Nils Bonfils},
      year={2025},
      eprint={2508.07454},
      booktitle={arXiv},
      primaryClass={cs.CY},
      howpublished={\url{https://arxiv.org/abs/2508.07454}}, 
}

@inproceedings{Alegri2025,
author = {Alegr\'{\i}a, Javiera and Bachmann, Ivana and Bustos-Jim\'{e}nez, Javier},
title = {Google Tag Manager and Its Privacy Issues},
year = {2025},
isbn = {978-3-032-06154-6},
publisher = {Springer-Verlag},
address = {Berlin, Heidelberg},
howpublished = {\url{https://doi.org/10.1007/978-3-032-06155-3_10}}, 
booktitle = {Security and Trust Management: 21st International Workshop, STM 2025, Toulouse, France, September 25–26, 2025, Proceedings},
pages = {173–189},
numpages = {17},
location = {Toulouse, France}
}

@article{Fouad2024,
  title = {The Devil is in the Details: Detection,  Measurement and Lawfulness of Server-Side Tracking on the Web},
  volume = {2024},
  ISSN = {2299-0984},
  howpublished = {\url{http://dx.doi.org/10.56553/popets-2024-0125}},
  DOI = {10.56553/popets-2024-0125},
  number = {4},
  journal = {Proceedings on Privacy Enhancing Technologies},
  publisher = {Privacy Enhancing Technologies Symposium Advisory Board},
  author = {Fouad,  Imane and Santos,  Cristiana and Laperdrix,  Pierre},
  year = {2024},
  month = oct,
  pages = {450–465}
}

@inproceedings{han2000,
author = {Han, Jiawei and Pei, Jian and Yin, Yiwen},
title = {Mining frequent patterns without candidate generation},
year = {2000},
isbn = {1581132174},
publisher = {Association for Computing Machinery},
address = {New York, NY, USA},
howpublished = {\url{https://doi.org/10.1145/342009.335372}},
doi = {10.1145/342009.335372},
booktitle = {Proceedings of the 2000 ACM SIGMOD International Conference on Management of Data},
pages = {1–12},
numpages = {12},
location = {Dallas, Texas, USA},
series = {SIGMOD '00}
}

@inproceedings{Nouwens2022,
author = {Nouwens, Midas and Bagge, Rolf and Kristensen, Janus Bager and Klokmose, Clemens Nylandsted},
title = {Consent-O-Matic: Automatically Answering Consent Pop-ups Using Adversarial Interoperability},
year = {2022},
isbn = {9781450391566},
publisher = {Association for Computing Machinery},
address = {New York, NY, USA},
howpublished = {\url{https://doi.org/10.1145/3491101.3519683}}, 
doi = {10.1145/3491101.3519683},
booktitle = {Extended Abstracts of the 2022 CHI Conference on Human Factors in Computing Systems},
articleno = {238},
numpages = {7},
location = {New Orleans, LA, USA},
series = {CHI EA '22}
}

@inproceedings{krumnow2022,
author = {Krumnow, Benjamin and Jonker, Hugo and Karsch, Stefan},
title = {How gullible are web measurement tools? a case study analysing and strengthening OpenWPM's reliability},
year = {2022},
isbn = {9781450395083},
publisher = {Association for Computing Machinery},
address = {New York, NY, USA},
howpublished = {\url{https://doi.org/10.1145/3555050.3569131}}, 
doi = {10.1145/3555050.3569131},
booktitle = {Proceedings of the 18th International Conference on Emerging Networking EXperiments and Technologies},
pages = {171–186},
numpages = {16},
location = {Roma, Italy},
series = {CoNEXT '22}
}

@inproceedings{Rack2023,
author = {Rack, Jeremy and Staicu, Cristian-Alexandru},
title = {Jack-in-the-box: An Empirical Study of JavaScript Bundling on the Web and its Security Implications},
year = {2023},
isbn = {9798400700507},
publisher = {Association for Computing Machinery},
address = {New York, NY, USA},
howpublished = {\url{https://doi.org/10.1145/3576915.3623140}},
doi = {10.1145/3576915.3623140},
booktitle = {Proceedings of the 2023 ACM SIGSAC Conference on Computer and Communications Security},
pages = {3198–3212},
numpages = {15},
location = {Copenhagen, Denmark},
series = {CCS '23}
}

@inproceedings{cahn2016,
author = {Cahn, Aaron and Alfeld, Scott and Barford, Paul and Muthukrishnan, S.},
title = {An Empirical Study of Web Cookies},
year = {2016},
isbn = {9781450341431},
publisher = {International World Wide Web Conferences Steering Committee},
address = {Republic and Canton of Geneva, CHE},
howpublished = {\url{https://doi.org/10.1145/2872427.2882991}}, 
booktitle = {Proceedings of the 25th International Conference on World Wide Web},
pages = {891–901},
numpages = {11},
location = {Montr\'{e}al, Qu\'{e}bec, Canada},
series = {WWW '16}
}

@misc{cookiegraph.artifact,
    author       = {Munir, Shaoor
                    and Siby, Sandra
                    and Iqbal, Umar
                    and Englehardt, Steven
                    and Shafiq, Zubair
                    and Troncoso, Carmela},
    title        = {{CookieGraph: Artifact Repository}},
    howpublished = {\url{https://github.com/cookiegraph/CookieGraph}},
    year         = {2023},
}

\appendix
\section{Open Science}
\label{sec:data_sharing}
To foster future research, we publish our analysis code, measurement data, data processing pipeline, and other supplementary information online at: \url{https://github.com/internet-sicherheit/From-Third-Party-to-First-Party-Measuring-and-Protecting-Against-Modern-Web-Tracking-Mechanisms}. A detailed description of how to use the developed tools to reproduce the analysis and rerun the crawling is available in the anonymous repository (see the \texttt{README.md} file). 

\section{Ethical Considerations}
\label{app:ethics}
As with all large-scale Web measurement studies, our study faces ethical challenges that must be considered.   
By running our crawler, we generate traffic and utilize resources, particularly those of the hosting parties for the analyzed sites (\eg energy), which could be used otherwise. Furthermore, the visited page might serve ads to our crawler, thereby consuming some of the advertiser's ad budget. Since each crawler visits each page only once, we assume these issues are minor and are widely accepted within the Web measurement community. %

\section{Overview of the Top 5 Clusters in the Ecosystem}
\label{app:ecosystem}
\begin{figure}[tb]
    \centering 
    \includegraphics[width=\columnwidth]{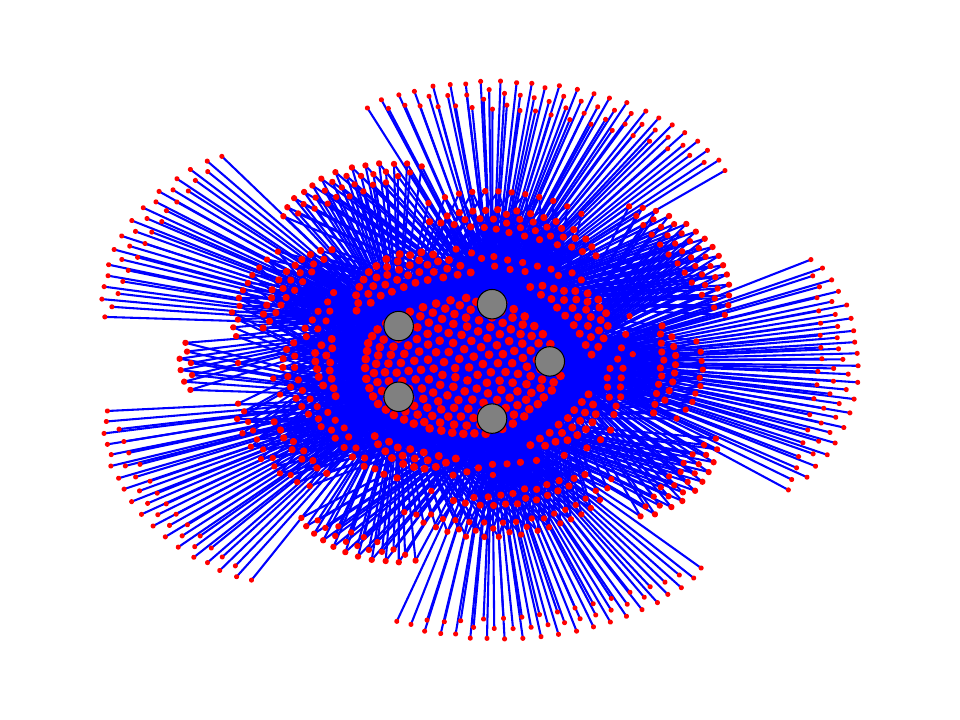}
     \caption{Network graph of the top 5 clusters by site data. Red dots represent cookies, and edges represent whether a cookie is used by a cluster (grey dots).}
    \label{fig:network_graph}
\end{figure}

\Cref{fig:network_graph} shows a graph of the top 5 clusters by site data. It is notable that we perform analysis on the complete graph but choose a subgraph for visualization. Therefore, we utilize the Python NetworkX library~\cite{hagberg_networkx_2008} to build the network graph. The figure shows the top five clusters and all the cookies associated with them. We define a node as a cookie (red dots) or cluster (gray dots). Edges (blue lines) between cookies and clusters represent whether a cookie is used by a script in a cluster. 

\section{Overview of the Heuristic Tests for Tracking Cookies}
\label{app:heuristicstests}
In \cref{tab:heuristic_test}, we provide an overview of the cookies we tested using the heuristic introduced in \cref{sec:cookies_identify}, and indicate at which step each cookie failed to meet the criteria for a tracking cookie. In these explorative tests, the heuristic did not flag any non-tracking cookies as tracking in the set of cookies we used to evaluate. Further, it did not flag the \texttt{s\_v\_web\_id} cookie (which is only valid for a session), as it is used to identify videos a user has seen, according to its description~\cite{cookie_svweb.2026}. Thus, we assume that our approach provides a lower bound. 

\begin{table*}[tb]
    \centering
    \resizebox{\textwidth}{!}{
        \centering
\begin{tabular}{llllccccc}
    \toprule
    \multicolumn{4}{c}{} & 
    \multicolumn{4}{c}{{\bfseries Criteria}} & 
    \multicolumn{1}{c}{} \\
    {\bfseries Cookie Name}
    &  {\bfseries Lifetime} 
    & {\bfseries Purpose}
    & {\bfseries Provider}
    & {\bfseries 1}
    & {\bfseries 2}
    & {\bfseries 3}
    & {\bfseries 4}
    & {\bfseries Result}
    \\
  
\midrule
\_ga & 2 years & Performance: used to distinguish unique users & Google &  \cmark & \cmark& \cmark&  \cmark & \cmark  \\

NID & 6 months & Targeting/Advertising: optimize targeted ad & Google & \cmark & \cmark & \cmark & \cmark & \cmark\\

SSID & 2 years & Targeting/Advertising: info on ads seen by the user & Google & \cmark&\cmark& \cmark&\xmark & \xmark \\

\_gat & session & Performance: bot detection & Google & \xmark &---& ---& ---& \xmark \\
\_gid & 24 hours & Performance: Used to distinguish users & Google &\cmark&\cmark &\cmark& \cmark& \cmark\\
\midrule

\_fbp & 90 days & Targeting/Advertising: track visits across websites & Meta & \cmark & \cmark & \cmark & \cmark & \cmark\\
xs & session& Targeting/Advertising: unique session ID & Meta & \xmark & --- & --- & --- & \xmark  \\
fr & 90 days & Targeting/Advertising: ad delivery or retargeting & Meta &\cmark & \cmark & \cmark& \cmark& \cmark\\
\midrule

ttwid & 1 year & Targeting/Advertising: track user content interaction & TikTok & \cmark & \cmark & \cmark & \cmark & \cmark\\
s\_v\_web\_id & session & Targeting/Advertising: identify videos seen by the user & TikTok & \xmark &---& ---& --- & \xmark \\

\midrule
mbox & 30 minutes & Functionality: Session ID, ID for browser   & Adobe& \cmark &\cmark& \cmark& \cmark& \cmark \\
 AMCV & 2 years & Targeting/Advertising: Unique identifier  & Adobe & \cmark & \cmark & \cmark & \cmark & \cmark\\
 s\_fid & 2 years & Performance: contains a unique id   & Adobe & \cmark &\cmark& \cmark& \cmark& \cmark \\
 s\_vi & 2 years & Performance: contains a unique ID and timestamp   & Adobe & \cmark &\cmark& \cmark& \cmark& \cmark \\
 demdex & 6 months &Targeting/Advertising: Identification, ID synchronization   & Adobe & \cmark &\cmark& \xmark& --- & \xmark \\

\bottomrule
\end{tabular}

    }
    \caption{Overview of the test to verify if the heuristic correctly identifies known first-party tracking cookies. Criteria 1: Longevity, 2: Length, 3: Uniqueness, 4:~Similarity.}
    \label{tab:heuristic_test}
\end{table*}

To validate the heuristic, we manually analyze the top 15 cookies identified by the heuristic to determine whether they are or could be used for user tracking; \Cref{tab:top_cookies_per_cluster} shows the top 10 of these.
For testing classification, we relied on sources such as Cookiepedia~\cite{Cookiepedia.2024}, Cookiedatabase~\cite{cookiedatabase.2026}, and general Internet searches.
Of the top ten cookies, six are directly related to tracking or analytics services and are therefore correctly classified.
For the other four (depicted with \specialCat ~in \cref{tab:top_cookies_per_cluster}), a more detailed analysis is needed, as they are not \emph{only} used for tracking. 
The cookie \texttt{\_abck} is primarily used by Akamai for bot detection, but it is also used for user tracking and analytics services~\cite{abck.cookie.2026,abck.cookie.comp.2026}.
\texttt{OptanonConsent} is used by OneTrust to store users' consent preferences. 
Various studies show that such consent management platform (CMP) providers can act as cross-site trackers, tracking users' choices across the Web~\cite{Santos.CMP.2021,Liu.Opt.out.2024,Rasaii.cmp.2025,Nouwens.cmp.2025}.
Amazon Web Services uses the cookie \texttt{AWSALB} primarily to enable load balancing. Yet, a primary use case of the cookie is to store session data locally on a specific server (here, AWS instance) rather than in a shared database. While technically not always linked to user tracking, it enables user tracking when scaling an application horizontally.
Due to an encrypted value that increases entropy and due to ``sticky sessions'' (\ie long expired times), the cookie passes the heuristic checks.
Finally, \texttt{PHPSESSID} is a configurable cookie commonly used by PHP frameworks to store session IDs, which are not a form of user tracking that we are interested in. However, when manually inspecting the cookies flagged by the heuristic, we noticed that the applications set the cookie's lifetime to ``infinity'' (\ie \texttt{9999-12-31 23:59:59 UTC}; the maximum value in the \texttt{YYYY-MM-DD} format). An infinite session can be used for user tracking.
We consider the last two cases as edge cases (probable false positives).

\begin{table*}[tb]
    \centering
    \centering
\begin{tabular}{clrrrl|cl}
\toprule
{\bfseries No.} 
& {\bfseries Cookie Name}
& {\bfseries Clusters}
& {\bfseries 1st Party}
& {\bfseries 3rd Party}
& {\bfseries Attributed to} 
& {\bfseries Tracking } 
& {\bfseries Purpose } 
\\
\midrule

1 & \_gcl\_au          & 5{,}102 & 1,417,269 & 54,447 & Google & \cmark & Analytics\\
2 & \_ga               & 4{,}907 & 1,365,268 & 49,366 & Google & \cmark & Performance\\
3 & \_abck             & 2{,}262 & 1,195,751 & 29,706 & Akamai & \specialCat & Bot detection\\
4 & \_\_smn\_fid       & 1{,}671 & 1,791 & 0 & Ladsp  & \cmark & Targeting\\ %
5 & AMCV               & 1{,}601 & 1,228,951& 33,951& Adobe  &\cmark & Targeting \\ 
6 & OptanonConsent     & 1{,}506 & 1,251,312 &22,862& OneTrust & \specialCat & Consent mgnt\\
7 & AWSALB             & 1{,}108 & 1,192,422 & 20,959& Amazon Web Services & \specialCat & Load balancing\\
8 & \_vwo\_uuid\_v2     & 1{,}077 & 1,262,926 &  28,539& Visual Website Optimizer & \cmark & Analytics \\
9 & \_fbp              & 972     & 1,203,112 &20,419 & Meta (Facebook) & \cmark & Targeting  \\
10 & PHPSESSID         & 860     & 1,201,277 & 18,874 & Unknown & \specialCat & Session mgnt\\

\bottomrule
\end{tabular}

    \caption{Top 10 cookies by their occurrence in the clusters. \specialCat~indicates that tracking is not the primary purpose of the cookie, but using it for tracking is possible.}
    \label{tab:top_cookies_per_cluster}
\end{table*}

To further validate the heuristic, we tested the distinct cookies identified in our clusters against Cookiepedia classifications. Of the 3,378 distinct cluster cookies, 1,829 (54\%) could not be classified by the database. 
About 262 (8\%) are classified as ``Targeting/Advertising''. 
By looking into the remaining 38\% of the cookies, we need to evaluate the heuristic. Those 38\% are distributed to ``Performance'' (805, 24\%), ``Strictly Necessary'' (266, 8\%), and ``Functionality'' (217, 6\%). Which leads to a ~14\% false-positive rate. As we learned from related work, some ``Performance'' cookies (\eg \texttt{\_ga}) can be used for tracking~\cite{cahn2016}.
An extended evaluation covering 100 sampled cookies is available in our repository (see Appendix~\ref{sec:data_sharing}).

\end{document}